\documentclass[twocolumn,showpacs,showkeys,preprintnumbers,amsmath,amssymb,prd,letterpaper,floatfix,superscriptaddress,10pt,aps,nofootinbib]{revtex4-1}
\usepackage[margin=1in,footskip=0.25in]{geometry}
\usepackage[utf8]{inputenc}
\usepackage{graphicx}
\usepackage{epsfig}
\usepackage{color, colortbl}
\usepackage{longtable}
\usepackage[breaklinks=true]{hyperref}
\usepackage{latexsym}
\usepackage{amsmath}
\usepackage{amssymb}
\usepackage{dsfont}
\usepackage{verbatim}
\usepackage{bm}
\usepackage{bbm}
\usepackage{placeins}
\usepackage[export]{adjustbox}
\usepackage{cleveref}
\usepackage{newunicodechar}
\usepackage{makecell}
\usepackage{braket}
\usepackage{ bbold }
\usepackage{multirow}
\usepackage{booktabs}
\usepackage{array}

\hypersetup{colorlinks, linkcolor = [rgb]{0,0.0,0.75}, citecolor = [rgb]{0,0.0,0.75}, urlcolor = [rgb]{0,0.0,0.75}}

\newcommand{\Nucleon}{N}

\def\MIT{Center for Theoretical Physics, Laboratory for Nuclear Science and Department of Physics, Massachusetts Institute of Technology, Cambridge, MA 02139, USA}
\def\RFWUB{Helmholtz-Institut f\"ur Strahlen- und Kernphysik, Rheinische Friedrich-Wilhelms-Universit\"at Bonn, Nu{\ss}allee 14-16, 53115 Bonn, Germany}
\def\CYI{Computation-based Science and Technology Research Center, The Cyprus Institute, 20 Kavafi Str., Nicosia 2121, Cyprus}
\def\UCY{Department of Physics, University of Cyprus, P.O. Box 20537, 1678 Nicosia, Cyprus}
\def\UTV{Dipartimento di Fisica, Università di Roma ``Tor Vergata'',  Via della Ricerca Scientifica 1, 00133 Rome, Italy}
\def\RIKEN{RIKEN BNL Research Center, Brookhaven National Laboratory, Upton, NY 11973, USA}

\def\UOA{Department of Physics, University of Arizona, Tucson, AZ 85721, USA}
\def\SBU{Department of Physics and Astronomy, Stony Brook University, Stony Brook, NY 11794, USA}
\def\BNL{Department of Physics, Brookhaven National Laboratory, Upton, NY 11973, USA}
\def\JLAB{Thomas Jefferson National Accelerator Facility, Newport News, VA 23606, USA}
\def\JSC{Forschungszentrum J\"ulich GmbH, J\"ulich Supercomputing Centre, 52425 J\"ulich, Germany}
\def\ODU{Department of Physics, Old Dominion University, Norfolk, VA 23529, USA}
\def\MAINZ{Institut f\"ur Kernphysik, Johannes Gutenberg-Universit\"at Mainz, 55099 Mainz, Germany}
\def\WUP{Faculty of Mathematics und Natural Sciences, University of Wuppertal
Wuppertal-42119, Germany}

\begin{document}

\author{Giorgio Silvi}
\email{giorgiosilvi@gmail.com}
\affiliation{\JSC}
\affiliation{\WUP}

\author{\mbox{Srijit Paul}}
\affiliation{\MAINZ}

\author{Constantia Alexandrou}
\affiliation{\UCY}
\affiliation{\CYI}

\author{Stefan Krieg}
\affiliation{\JSC}
\affiliation{\RFWUB}

\author{Luka Leskovec}
\affiliation{\JLAB}
\affiliation{{\ODU}}

\author{Stefan Meinel}
\affiliation{\UOA}

\author{John Negele}
\affiliation{\MIT}

\author{Marcus Petschlies}
\affiliation{\RFWUB}

\author{Andrew Pochinsky}
\affiliation{\MIT}

\author{Gumaro Rendon}
\affiliation{\BNL}

\author{Sergey Syritsyn}
\affiliation{\SBU}
\affiliation{\RIKEN}

\author{Antonino Todaro}
\affiliation{\UCY}
\affiliation{\UTV}
\affiliation{\WUP}

\title{\texorpdfstring{$\bm{P}$}{P}-wave nucleon-pion scattering amplitude in the \texorpdfstring{$\bm{\Delta (1232)}$}{Delta(1232)} channel \\ from lattice QCD}
\date{\today}

\begin{abstract}
  We determine the $\Delta(1232)$ resonance parameters using lattice QCD and the L{\"u}scher method.
  The resonance occurs in elastic pion-nucleon scattering with $J^P=3/2^+$ in the isospin $I = 3/2$, $P$-wave channel.
  Our calculation is performed with $N_f=2+1$ flavors of clover fermions on a lattice with $L\approx 2.8$ fm. The pion and nucleon masses
  are $m_\pi =255.4(1.6)$ MeV and $m_N=1073(5)$ MeV, respectively, and the strong decay channel $\Delta \rightarrow \pi N$ is found to be above the threshold.
  To thoroughly map out the energy-dependence of the nucleon-pion scattering amplitude, we compute the spectra in all relevant irreducible
  representations of the lattice symmetry groups for total momenta up to $\vec{P}=\frac{2\pi}{L}(1,1,1)$, including irreps that mix $S$ and $P$ waves.
  We perform global fits of the amplitude parameters to up to 21 energy levels, using a Breit-Wigner model for the $P$-wave phase shift and the effective-range expansion for the $S$-wave phase shift.   From the location of the pole in the $P$-wave scattering amplitude, we obtain the resonance mass $m_\Delta=1378(7)(9)$ MeV and the coupling $g_{\Delta\text{-}\pi N}=23.8(2.7)(0.9)$.
\end{abstract}

\maketitle

\section{Introduction}\label{sec_introduction}

The $\Delta(1232)$ (in the following denoted as $\Delta$) is the lowest-lying baryon resonance, typically produced when energetic photons, neutrinos, or pions hit a nucleon \cite{alvarez2018nustec}. While these three processes differ immensely, they have the two-particle nucleon-pion scattering amplitude in common. The scattering amplitude in which the $\Delta$ appears as an enhancement in the $P$-wave with $J^P=\frac{3}{2}^+$ and $I=\frac{3}{2}$, often also referred to as the $P_{33}$ amplitude, where the notation means $l_{2I\,2J}$. For energies near the $\Delta$ mass, this amplitude is nearly completely elastic \cite{roper1965energy,meissner2001progress}.

Modern determinations of the $\Delta$ resonance parameters are typically performed using data from experiments such as CLAS12 at JLab and MAMI-A1 in Mainz. While the results for the pole location differ slightly from the Breit-Wigner parameters \cite{Djukanovic:2007bw}, the $\Delta$ is generally found to have a mass of approximately $1230$ MeV and a decay width of approximately $100$ MeV \cite{PDG2020,shrestha2012multichannel}.

Phenomenological studies of the $\Delta$ have been performed using quark models, chiral perturbation theory and related effective field theories, and the S-matrix approach. From the quark-model point of view, many baryons remain elusive, but the $\Delta$ mass is reproduced quite well \cite{Hemmert:1994ky,Capstick:2000qj,Tiator:2010rp}. Chiral perturbation theory and related effective field theories have shown great success in determining low-energy scattering parameters and $\pi N$ scattering amplitudes \cite{Mojzis:1997tu,Hemmert:1997ye,Fettes:1998ud,Long:2009wq}; an extensive review can be found in Ref.~\cite{Pascalutsa:2006up}. Analyses of the large experimental data sets using amplitude
models based on S-matrix principles were performed in Refs.~\cite{Matsinos:2006sw,Arndt:2006bf,Beck:2016hcy}.

First-principles computations of $\Delta$ properties can be done using lattice QCD. The $\Delta$ mass, assuming a stable $\Delta$, was studied in Refs.~\cite{Basak:2007kj,Edwards:2011jj,Dudek:2012ag,Gattringer:2008vj,Engel:2010my,Engel:2013ig}. However, for quark masses corresponding to pion masses below a certain value, the $\Delta$ is an unstable hadron, and its mass and decay width must be determined from the appropriate $N\pi$ scattering amplitudes. While the use of Euclidean time in lattice QCD prevents direct computations of infinite-volume scattering amplitudes \cite{maiani1990final}, L{\"u}scher showed how the finite-volume energy spectrum of a two-body system interacting through an elastic short-range interaction is related to the infinite-volume scattering amplitudes \cite{luescher1986volume,luscher1990calculate,luscher1991two}. The decades following L{\"u}scher's seminal work witnessed further development of the theoretical framework to moving frames \cite{rummukainen1995resonance,feng2011new}, unequal masses \cite{davoudi2011improving,fu2012rummukainen,leskovec2012scattering}, and arbitrary spin \cite{briceno2014two}. These methods have been applied to many systems in the meson sector and are reviewed in Ref.~\cite{Briceno:2017max}. For the nucleon-pion scattering only a handful of studies have been done in the $N\pi$ channel \cite{Mohler:2012nh,lang2013scattering,alexandrou2013determination,verduci2014pion,alexandrou2016study,lang2017pion,andersen2018elastic,andersen20193,Mei_ner_2011}.

In the following, we report a new lattice-QCD study of elastic $N\pi$ scattering in the $\Delta$ resonance channel using the L{\"u}scher method. Our calculation is performed using  $N_f=2+1$ flavors of clover fermions at a pion mass of $m_\pi =255.4(1.6)$ MeV, on a lattice with $L\approx 2.8$ fm. We obtain detailed results for the energy-dependence of the scattering amplitude by analyzing multiple moving frames. From the amplitude's pole position, we determine the $\Delta$ mass, decay width, and its coupling to the $N\pi$ channel. Preliminary results were previously shown in Ref.~\cite{paul2018towards}. The computations presented here also constitute the first step toward a future calculation of $N\to N\pi$ electroweak transition matrix elements using the formalism of Refs.~\cite{Briceno:2014uqa,Briceno:2015csa}.

The paper is organized as follows: in Sec.~\ref{sec_gauge_ensemble} the details of the lattice gauge-field ensemble are presented. Section \ref{sec_int_operators} describes the interpolating operators and the method used to project to definite irreducible representations of the lattice symmetry groups. The Wick contractions yielding the two-point correlation functions for the $\Delta-N\pi$ system are discussed in Sec.~\ref{sec_wick_contr}. In Sec.~\ref{sec_spectra_results}, the results of the spectra analysis are presented. The relevant finite-volume quantization conditions are discussed in Sec.~\ref{sec_luscher_analysis}. The $K$-matrix parametrizations employed for the scattering amplitudes and our results for the amplitude parameters are presented in Sec.~\ref{sec_result}. We conclude in Sec.~\ref{sec_conclusions}.


\section{Gauge Ensemble}\label{sec_gauge_ensemble}

\begin{table}

  \begin{tabular}{|>{\columncolor[gray]{.9}[\tabcolsep]}c|c|}

    \hline
    $N_s^3\times N_t$ & $24^3\times 48$ \cr
    $\beta$           & $3.31$  \cr
    $a m_{u,d}$       & $-0.09530$ \cr
    $a m_s$           & $-0.040$  \cr
    $a\:[\rm fm]$     & $ 0.1163(4)$ \cr
    $L \:[\rm fm]$    & $2.791(9)$ \cr
    $m_\pi$ [MeV]     & $255.4(1.6)$ \cr
    $m_\pi L$         & $3.61(2)$ \cr
    $N_{config}$      & $600$ \cr
    $N_{meas}$        & $9600$ \cr
    \hline
  \end{tabular}

  \caption{Parameters of the lattice gauge-field ensemble.}
  \label{tab:lattice}
\end{table}

We use a lattice gauge-field ensemble generated with the setup of the Budapest-Marseille-Wuppertal collaboration \cite{Durr:2010aw}, with parameters given in Table \ref{tab:lattice}. The ensemble has been used previously in Ref.~\cite{green2014nucleon}. The gluon action is the tree-level improved Symanzik action \cite{luscher1985computation}, while the fermion action is a tree-level clover-improved Wilson action \cite{sheikholeslami1985improved} with two levels of HEX smearing of the gauge links \cite{Durr:2010aw}. We analyze 600 gauge configurations and compute the correlation functions for 16 source positions on each configuration, resulting in a total of 9600 measurements.

When considering the $N\pi$ system in the rest frame only, the spatial lattice size of $L\approx 2.8$ fm (with periodic boundary conditions) results in a rather sparse energy spectrum across the elastic region.
Between the $N\pi$ and $N\pi\pi$ thresholds there are few energy points available to constrain the phase shift we aim to determine.
A straightforward way to gain additional points would be to add a spatially larger ensemble, but this is computationally quite expensive.
A more efficient approach employed here is using also moving frames \cite{rummukainen1995resonance,gockeler2012scattering,kim2005finite} on the same ensemble, where the Lorentz boost contracts the box, resulting in different effective values of the spatial length along the boost direction \cite{leskovec2012scattering}.

\section{Interpolating Operators}\label{sec_int_operators}

\begin{table*}
  \centering

  \begin{tabular}{| >{\columncolor[gray]{.9}[\tabcolsep]}c | c | c | c | c |}
    \hline
    $\frac{L}{2\pi}\vec{P}  $              & $(0,0,0)$                                        & $(0,0,1)$                                                                  & $(0,1,1)$                               & $(1,1,1)$                                                                          \\
    \hline
    Group $LG$                             & $O_{h}^{(D)}$                                    & $C_{4v}^{(D)}$                                                             & $C_{2v}^{(D)}$                          & $C_{3v}^{(D)}$                                                                     \\
    \hline

    \Gape[0pt][2pt]{\makecell{Axis and planes                                                                                                                                                                                                                                                             \\ of symmetry}}& \includegraphics[valign=m,width=0.13\textwidth]{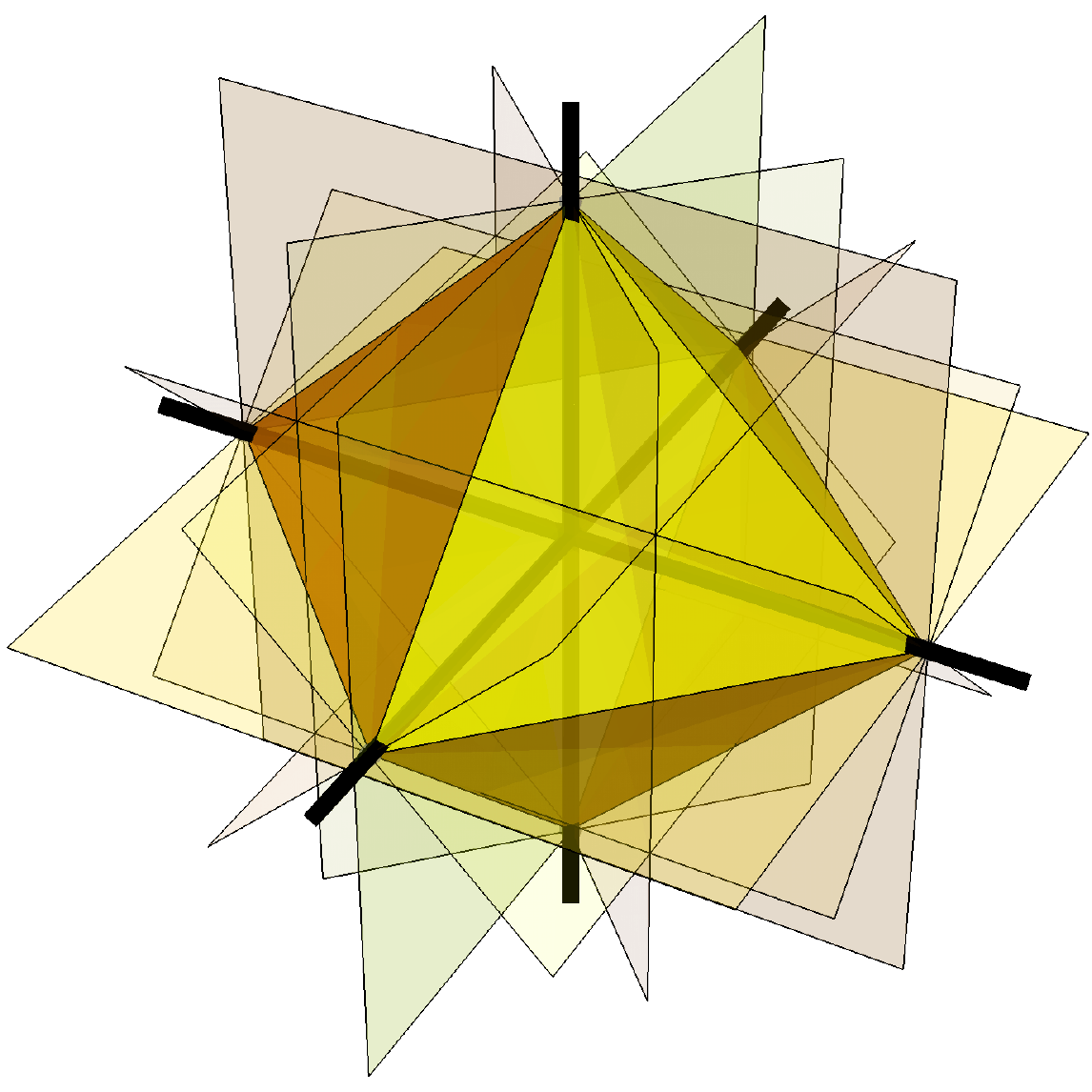} & \includegraphics[valign=m,width=0.13\textwidth]{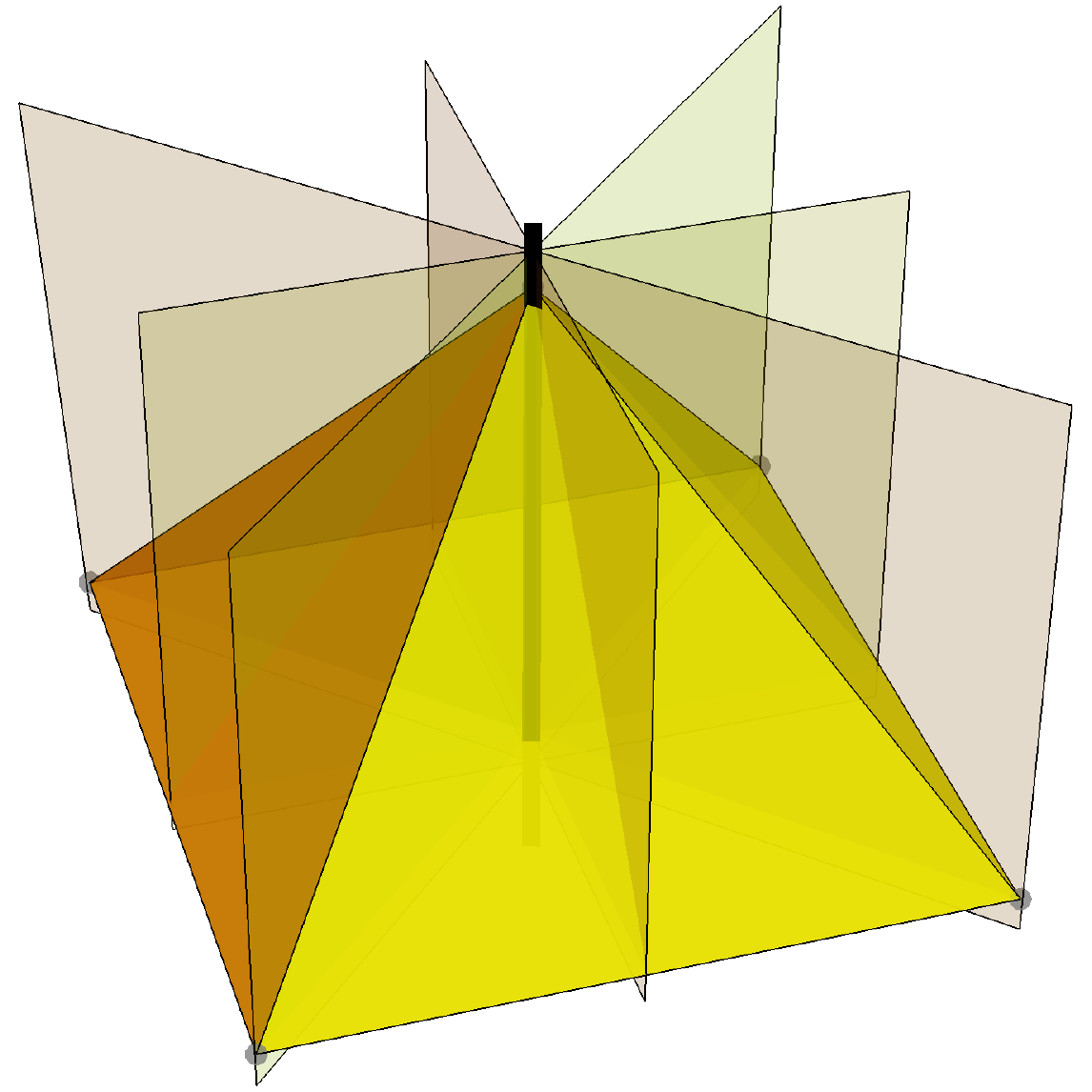} &\includegraphics[valign=m,width=0.13\textwidth]{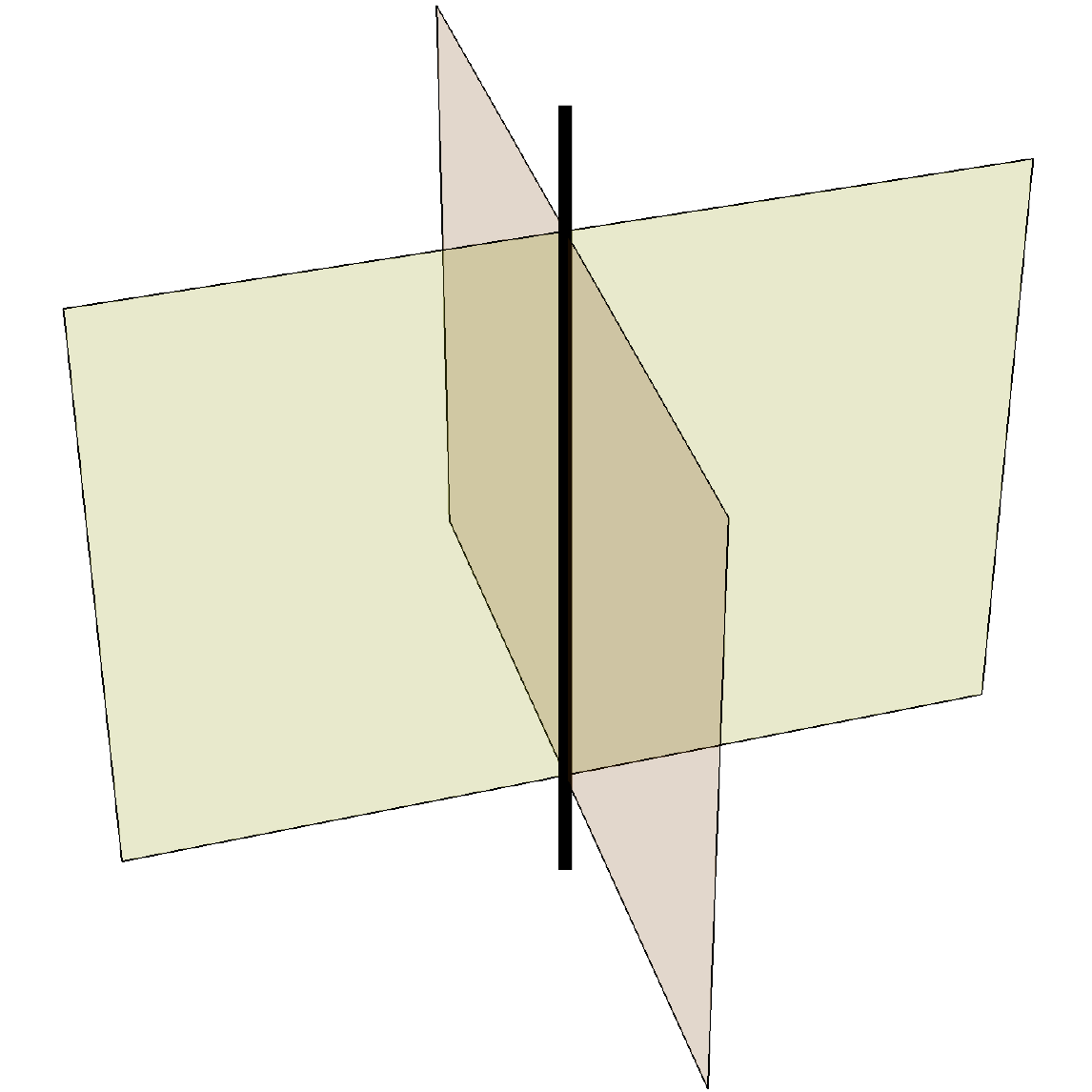}       & \includegraphics[valign=m,width=0.13\textwidth]{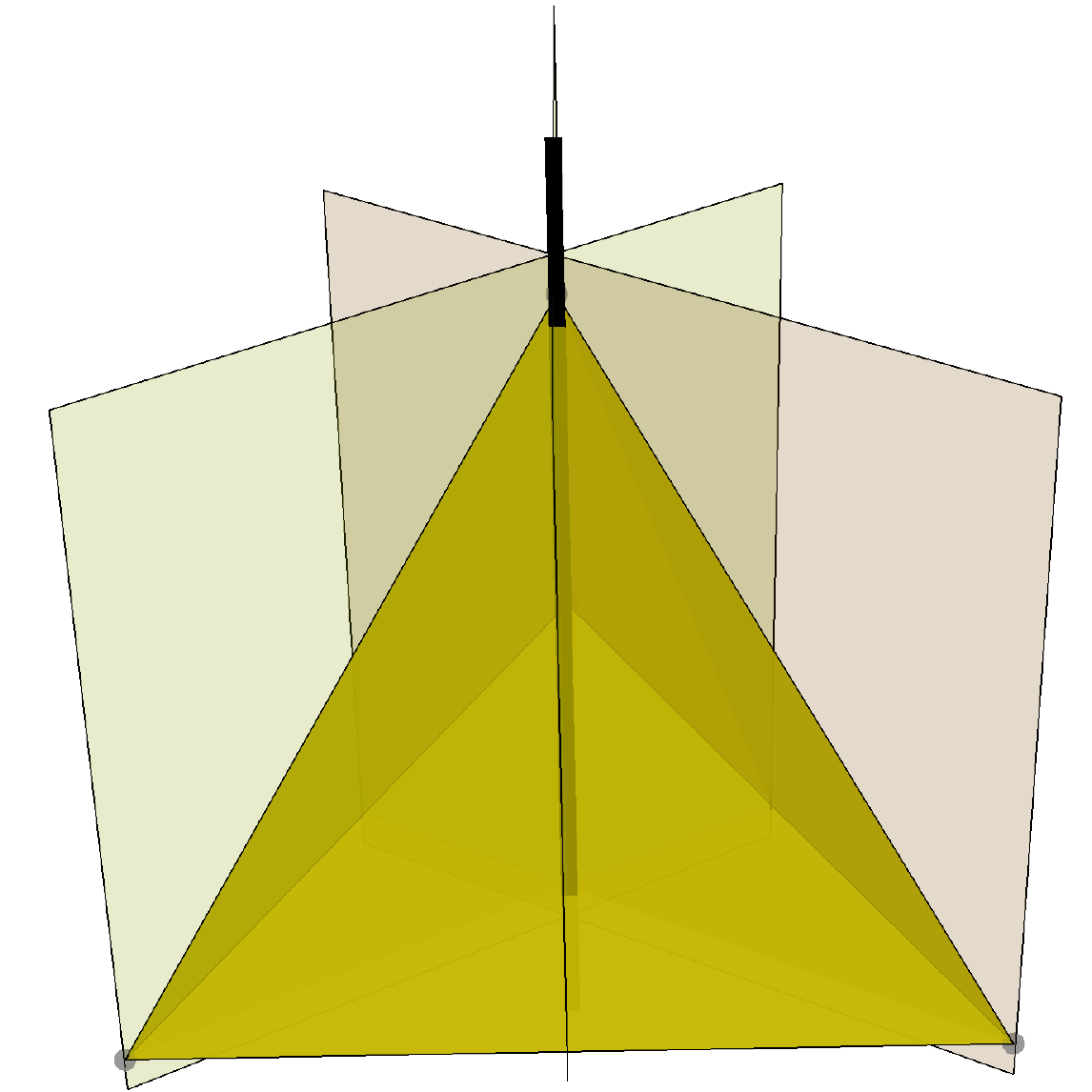}   \\
    \hline
    $g_{LG}$                               & 96                                               & 16                                                                         & 8                                       & 12                                                                                 \\
    \hline
    $\Lambda(J^P):\pi ($ $0^-$ $) $        & $A_{1u}($ $0^-$ $,4^-,...)$                      & $A_{2}($ $0$ $,1,...)$                                                     & $A_{2}($ $0$ $,1,...)$                  & $A_{2}($ $0$ $,1,...)$                                                             \\
    \hline
    $\Lambda(J^P):N ($ $\frac{1}{2}^+$ $)$ & $G_{1g}($ $ \frac{1}{2}^+$ $,\frac{7}{2}^+,...)$ & $G_{1}($ $ \frac{1}{2}$ $,\frac{3}{2},...)$                                & $G($ $ \frac{1}{2}$ $,\frac{3}{2},...)$ & $G($ $\frac{1}{2}$ $,\frac{3}{2},...)$                                             \\
    \hline
    $\Lambda(J^P):\Delta (\frac{3}{2}^+)$  & $H_{g}(\frac{3}{2}^+,\frac{5}{2}^+,...)$         & $G_1(\frac{1}{2},\frac{3}{2},...) \oplus G_2(\frac{3}{2},\frac{5}{2},...)$ & $(2)G(\frac{1}{2},\frac{3}{2},...)$     & \makecell{$G(\frac{1}{2},\frac{3}{2},...) \oplus F_1(\frac{3}{2},\frac{5}{2},...)$ \\  $\oplus F_2(\frac{3}{2},\frac{5}{2},...)$} \\
    \hline
  \end{tabular}
  \caption{Choices of total momenta $\vec{P}$, along with the Little Groups $LG$, irreducible representations $\Lambda$ of relevant hadrons and their angular momentum content $J^P$.
    The multi-hadron $N\pi$ operators have the same irreps as the single-hadron $\Delta$ operators. From left to right the subduction of irreps in moving frames. The label $``(2)"$ for irrep $G$ in group $C_{2v}^D$ indicates the double occurrence of the irrep from the subduction; to differentiate this irrep from the homonymous of group $C_{3v}^D$ we keep the additional label $``(2)"$ throughout the paper. Images credit \cite{pointgroups}.}
  \label{tab:plan}
\end{table*}

We use local single-hadron and nonlocal multi-hadron interpolating operators, both necessary for a complete determination of the resonance properties \cite{wilson2015coupled}.
For the single-hadron $\Delta$ operators with $I=3/2,I_3=+3/2$ (corresponding to the $\Delta^{++}$) we include two choices,
\begin{equation}
  \label{eq:delta_op}
  \begin{split}
    \Delta_{\alpha i}^{(1)} (\vec{p})= \sum_{\vec{x}} \epsilon_{abc}
    ( u_{a}(\vec{x}))_\alpha ( u_b^T(\vec{x})C \gamma_i u_c(\vec{x}))\,  e^{i\vec{p}\cdot\vec{x}}, \\
    \Delta_{\alpha i}^{(2)} (\vec{p})= \sum_{\vec{x}} \epsilon_{abc}
    ( u_{a}(\vec{x}))_\alpha ( u_b^T(\vec{x})C \gamma_i \gamma_0 u_c(\vec{x}))\,  e^{i\vec{p}\cdot\vec{x}}.
  \end{split}
\end{equation}
The two-hadron interpolators with the same quantum numbers are obtained from products of the form
\begin{equation}
  \Nucleon_\alpha^{(1,2)} (\vec{p}_{1})\,\pi(\vec{p}_{2})
\end{equation}
as explained in more detail below. The pion interpolator ($I=1,I_3=+1$) is given by
\begin{equation}
  \label{eq:pion_op}
  \pi^+(\vec{p})=\sum_x \bar{d}(\vec{x}) \gamma_5 u(\vec{x}) e^{i\vec{p}\cdot\vec{x}},
\end{equation}
and for the nucleon ($ I=1/2,I_3=+1/2$) we again include two choices,
\begin{equation}
  \label{eq:nucleon_op}
  \begin{split}
    \Nucleon_\alpha^{(1)} (\vec{p})= \sum_{\vec{x}} \epsilon_{abc}
    ( u_{a}(\vec{x}))_\alpha ( u_b^T(\vec{x})C \gamma_5 d_c(\vec{x})) \,  e^{i\vec{p}\cdot\vec{x}}, \\
    \Nucleon_\alpha^{(2)} (\vec{p})= \sum_{\vec{x}} \epsilon_{abc}
    (u_{a}(\vec{x}))_\alpha ( u_b^T(\vec{x})C\gamma_0\gamma_5 d_c(\vec{x})) \, e^{i\vec{p}\cdot\vec{x}}\,.
  \end{split}
\end{equation}

To correctly identify the angular momentum in the reduced symmetry of the cubic box, we project the operators to the irreducible representations (irreps) that belong to the symmetry groups of the finite volume.
Instead of the infinitely many possible irreducible representations $J^P$ of the continuum, on the lattice, there are only a finite number of possible irreps $\Lambda$.
Thus each lattice irrep in principle contains infinitely many values of the continuum spin $J$.
Each irrep belongs to a Little Group $LG(\vec{P})$ describing the underlying symmetry of the finite spatial volume contracted in the direction of the boost vector $\vec{P}$, i.e., the total momentum of the $N\pi$ system.

In the moving frames considered here, the symmetries are reduced to the groups $C_{4v}, C_{2v}, C_{3v}$ (see Table \ref{tab:plan}).
The degree of symmetry is mirrored by the group's order $g_{LG(\vec{P})}$, which corresponds to the number of transformation elements (rotations and inversions) belonging to the group.
In particular, half-integer spin is best described by the double-cover of symmetry groups (labeled $ D $), which introduce the $2\pi$ rotation as a new element of the group, effectively doubling the elements of the original group \cite{altmann2005rotations}.
Additionally, a clear parity identification is lost in the moving frames,
where the subduction mixes parities in the same irrep \cite{edwards2011excited}.
The list of chosen total momenta, symmetry groups, and irreps for the hadrons used in this work can be found in Table \ref{tab:plan}.

\begin{table*}
  \centering
  \begin{tabular}{ |c|c|c|c|c|c|c|  }
    \hline
    \rowcolor[gray]{.9}[\tabcolsep]
    \makecell{$\frac{L}{2\pi}\vec{P}_{ref}$                                                                                                                  \\ $[N_{dir}]$ } & Group $LG$ & Irrep $\Lambda$ & Rows & Ang.~mom.~content & Operator structure  & Number of operators  \\
    \hline
    (0,0,0) & $O_h^D$    & $G_{1u}$ & 2 & $J=1/2,7/2,...$ & $\mathcal{N}\pi$ with $|\vec{p}_1|=|\vec{p}_2|=0$                                           & 1  \\
    $[1]$   &            &          &   &                 & $\mathcal{N}\pi$ with $|\vec{p}_1|=|\vec{p}_2|=\frac{2\pi}{L}$                              & 2  \\
    \cline{3-7}
            &            & $H_{g}$  & 4 & $J=3/2,5/2,...$ & $\Delta^{(1,2)}(\vec{P})$                                                                   & 2  \\
            &            &          &   &                 & $\mathcal{N}\pi$ with $|\vec{p}_1|=|\vec{p}_2|=\frac{2\pi}{L}$                              & 2  \\
    \hline
    (0,0,1) & $C_{4v}^D$ & $G_1$    & 2 & $J=1/2,3/2,...$ & $\Delta^{(1,2)}(\vec{P})$                                                                   & 8  \\
    $[3]$   &            &          &   &                 & $\mathcal{N}\pi$ with $|\vec{p}_1|=0$ and $|\vec{p}_2|=\frac{2\pi}{L}$                      & 2  \\
            &            &          &   &                 & $\mathcal{N}\pi$ with $|\vec{p}_1|=\frac{2\pi}{L}$ and $|\vec{p}_2|=0$                      & 2  \\
            &            &          &   &                 & $\mathcal{N}\pi$ with $|\vec{p}_1|=\frac{2\pi}{L}$ and $|\vec{p}_2|=\sqrt{2}\frac{2\pi}{L}$ & 4  \\
            &            &          &   &                 & $\mathcal{N}\pi$ with $|\vec{p}_1|=\sqrt{2}\frac{2\pi}{L}$ and $|\vec{p}_2|=\frac{2\pi}{L}$ & 4  \\
    \cline{3-7}
            &            & $G_2$    & 2 & $J=3/2,5/2,...$ & $\Delta^{(1,2)}(\vec{P})$                                                                   & 4  \\
            &            &          &   &                 & $\mathcal{N}\pi$ with $|\vec{p}_1|=\sqrt{2}\frac{2\pi}{L}$ and $|\vec{p}_2|=\frac{2\pi}{L}$ & 4  \\
            &            &          &   &                 & $\mathcal{N}\pi$ with $|\vec{p}_1|=\frac{2\pi}{L}$ and $|\vec{p}_2|=\sqrt{2}\frac{2\pi}{L}$ & 4  \\
    \hline
    (0,1,1) & $C_{2v}^D$ & $(2)G$   & 2 & $J=1/2,3/2,...$ & $\Delta^{(1,2)}(\vec{P})$                                                                   & 12 \\
    $[6]$   &            &          &   &                 & $\mathcal{N}\pi$ with $|\vec{p}_1|=0$ and $|\vec{p}_2|=\sqrt{2}\frac{2\pi}{L}$              & 2  \\
            &            &          &   &                 & $\mathcal{N}\pi$ with $|\vec{p}_1|=\sqrt{2}\frac{2\pi}{L}$ and $|\vec{p}_2|=0$              & 2  \\
            &            &          &   &                 & $\mathcal{N}\pi$ with $|\vec{p}_1|=|\vec{p}_2|=\frac{2\pi}{L}$                              & 4  \\
    \hline
    (1,1,1) & $C_{3v}^D$ & $G$      & 2 & $J=1/2,3/2,...$ & $\Delta^{(1,2)}(\vec{P})$                                                                   & 8  \\
    $[4]$   &            &          &   &                 & $\mathcal{N}\pi$ with $|\vec{p}_1|=0$ and $|\vec{p}_2|=\sqrt{3}\frac{2\pi}{L}$              & 2  \\
            &            &          &   &                 & $\mathcal{N}\pi$ with $|\vec{p}_1|=\sqrt{3}\frac{2\pi}{L}$ and $|\vec{p}_2|=0$              & 2  \\
            &            &          &   &                 & $\mathcal{N}\pi$ with $|\vec{p}_1|=\frac{2\pi}{L}$ and $|\vec{p}_2|=\sqrt{2}\frac{2\pi}{L}$ & 4  \\
            &            &          &   &                 & $\mathcal{N}\pi$ with $|\vec{p}_1|=\sqrt{2}\frac{2\pi}{L}$ and $|\vec{p}_2|=\frac{2\pi}{L}$ & 4  \\
    \cline{3-7}
            &            & $F_1$    & 1 & $J=3/2,5/2....$ & $\Delta^{(1,2)}(\vec{P})$                                                                   & 4  \\
            &            &          &   &                 & $\mathcal{N}\pi$ with $|\vec{p}_1|=\frac{2\pi}{L}$ and $|\vec{p}_2|=\sqrt{2}\frac{2\pi}{L}$ & 2  \\
            &            &          &   &                 & $\mathcal{N}\pi$ with $|\vec{p}_1|=\sqrt{2}\frac{2\pi}{L}$ and $|\vec{p}_2|=\frac{2\pi}{L}$ & 2  \\
    \cline{3-7}
            &            & $F_2$    & 1 & $J=3/2,5/2,...$ & $\Delta^{(1,2)}(\vec{P})$                                                                   & 4  \\
            &            &          &   &                 & $\mathcal{N}\pi$ with $|\vec{p}_1|=\frac{2\pi}{L}$ and $|\vec{p}_2|=\sqrt{2}\frac{2\pi}{L}$ & 2  \\
            &            &          &   &                 & $\mathcal{N}\pi$ with $|\vec{p}_1|=\sqrt{2}\frac{2\pi}{L}$ and $|\vec{p}_2|=\frac{2\pi}{L}$ & 2  \\

    \hline
  \end{tabular}
  \caption{List of projected single-hadron ($\Delta$) and multi-hadron ($\mathcal{N}\pi$) operators for all irreps. In the construction of the multi-hadron operators, we use optimized nucleon operators $\mathcal{N}$ that are linear combinations of $\Nucleon^{(1)}$ and $\Nucleon^{(2)}$, as defined in Eq.~(\ref{eq:nucleonoptimized}). \label{tab:all_irreps}}
\end{table*}

To project the single-hadron operators to a definite irrep $\Lambda$ and row $r$, we make use of the formula \cite{basak2007lattice,prelovsek2017lattice,basak2005group,morningstar2013extended,bernard2008resonance}:
\begin{align}
  \label{eq:singlehadron}
  O^{\Lambda,r,i}(\vec{P})= \frac{d_\Lambda}{g_{LG(\vec{P})}}\sum_{R \in LG(\vec{P})} \Gamma_{r,r}^\Lambda (R)\: W (R)^{-1} O(\vec{P}),
\end{align}
where $d_\Lambda$ is the dimension of the irrep $\Lambda$ and $\Gamma^\Lambda$
are the representation matrices belonging to the irrep $\Lambda$. The matrices $W (R)^{-1}$ correspond to the matrices appearing in the right-hand sides
of Eqs.~(\ref{pion_transf}), (\ref{nucleon_transf}), or (\ref{delta_transf}). Here we denote the elements of the little group generically as $R$, even though in the rest frame they include the inversion in addition to the lattice rotations. The index $i$ labels the embedding into the irrep and replaces any free Dirac/Lorentz indices appearing on the right-hand side of Eq.~(\ref{eq:singlehadron}).

The analogous projection formula for the two-hadron operators is
\begin{align}
  \label{eq:multihadron}
  O_{N\pi}^{\Lambda,r,i}(\vec{P}) & = \frac{d_\Lambda}{g_{LG(\vec{P})}}\sum_{R \in LG(\vec{P})} \sum_{\vec{p}} \Gamma_{r,r}^\Lambda (R)  \nonumber \\
                                  & \times W_N^{-1} (R)  N(R\vec{p}) W_\pi (R)^{-1}\pi(\vec{P}-R\vec{p}).
\end{align}
Representation matrices for irreps in the rest frame are found in \cite{johnson1982angular,bernard2008resonance} and for the moving frames are provided in \cite{morningstar2013extended}.
In Eq.~(\ref{eq:multihadron}), given a total momentum $\vec{P}$, the sum over internal momenta is constrained by the magnitudes $|\vec{p}_{1}|=|R \vec{p}|=|\vec{p}|$ and $|\vec{p}_{2}|=|\vec{P}-R\vec{p}|$.
The structure of the projected operators $\Delta$ and $N\pi$ for all irreps is listed in Table \ref{tab:all_irreps}.

In general, both Eqs.~(\ref{eq:singlehadron}) and (\ref{eq:multihadron}) produce for each row $r$ of irrep $\Lambda$ multiple operator embeddings
(identified by the label $i$) that are not guaranteed to be independent. We therefore perform the following three steps to arrive at our final set of operators \cite{basak2005group}:
\begin{enumerate}
  \item [(i)] Construct all possible operators using Eqs.~(\ref{eq:singlehadron}) and (\ref{eq:multihadron}) for $r=1$ only.
  \item [(ii)] Reduce the sets of operators obtained in this way to linearly independent sets.
  \item [(iii)] Construct the other rows $r$ for these linearly independent sets of operators.
\end{enumerate}
The operators obtained in step (i) have the generic form
\begin{equation}
  O^{\Lambda,1,i}(\vec{P})=\sum_j c^{\Lambda,1}_{ij}O^j(\vec{P}).
\end{equation}
Using Gaussian elimination we obtain a smaller matrix $c^{\Lambda,1}_{nj}$ such that the linearly independent operators constructed in step (ii) have the form
\begin{equation}
  O^{\Lambda,1,n}(\vec{P})=\sum_j c^{\Lambda,1}_{nj}O^j(\vec{P}).
\end{equation}
The number of independent operators (corresponding to the range of the index $n$) is equal to \cite{johnson1982angular,cotton2003chemical}
\begin{equation}
  \label{occ}
  \frac{1}{g_{LG(\vec{P})}}\sum_{R \in LG(\vec{P})} \chi^{\Gamma^\Lambda}(R)\chi^{W}(R),
\end{equation}
where the characters $\chi^{\Gamma^\Lambda}(R)$ and $\chi^{W}(R)$ are equal to the traces of the representation matrices $\Gamma^\Lambda$ and the transformation matrices $W(R)$.

In step (iii), to construct the other rows $r>1$ we use
\begin{align}
  O^{\Lambda,r,n}(\vec{P})= & \sum_j c^{\Lambda,1}_{nj} \frac{d_\Lambda}{g_{LG(\vec{P})}} \nonumber                                    \\
                            & \times \sum_{R \in LG(\vec{P})} \Gamma_{r,1}^\Lambda (R) \, \mathsf{R}\, O^j(\vec{P})\, \mathsf{R}^{-1},
\end{align}
where the rotations/inversions $\mathsf{R}\, O^j(\vec{P})\, \mathsf{R}^{-1}$ are performed as in Eqs.~(\ref{eq:singlehadron}) and (\ref{eq:multihadron}), depending on the structure of $O^j(\vec{P})$.

Also, to increase statistics, multiple directions of $\vec{P}$ at fixed $|\vec{P}|$ are used (see Table \ref{tab:all_irreps}).
For every moving frame, we first perform the irrep projections for a reference momentum $\vec{P}_{ref}$ and then rotate the projected operators to the new momentum direction.
Generating operators initially from a reference momentum and $r=1$ only facilitates the identification of equivalent operators embeddings that can later be
averaged over different rows of the same irrep $\Lambda$ (which is possible due to the great orthogonality theorem \cite{cotton2003chemical}) and momentum direction of equal $|\vec{P}|$.
In the following, the label $r$ for the row will be dropped.

\section{Wick contractions}\label{sec_wick_contr}
\begin{figure}
  \centering
  \includegraphics[width=0.9\columnwidth]{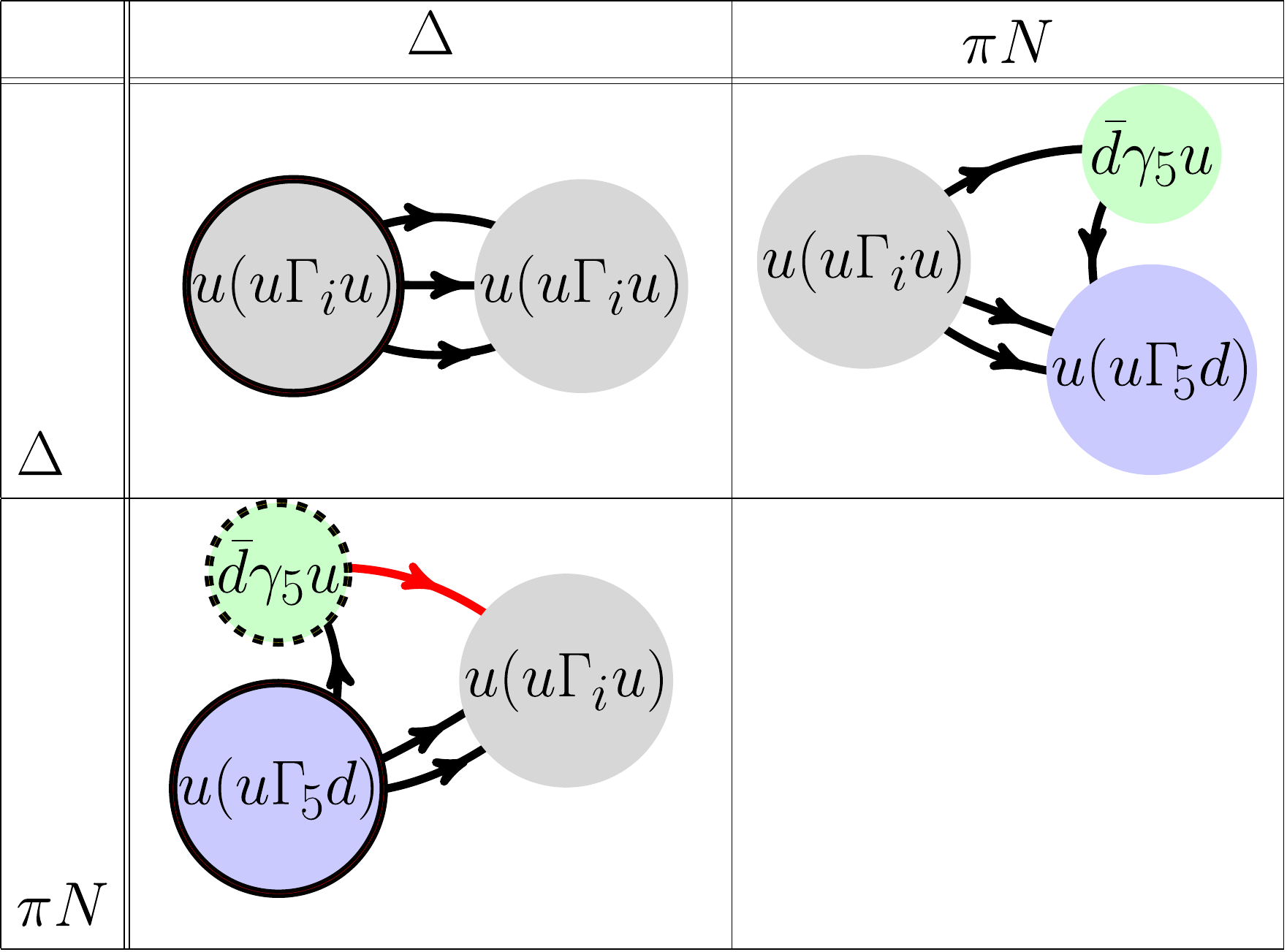}%
  \hfill
  \includegraphics[width=0.9\columnwidth]{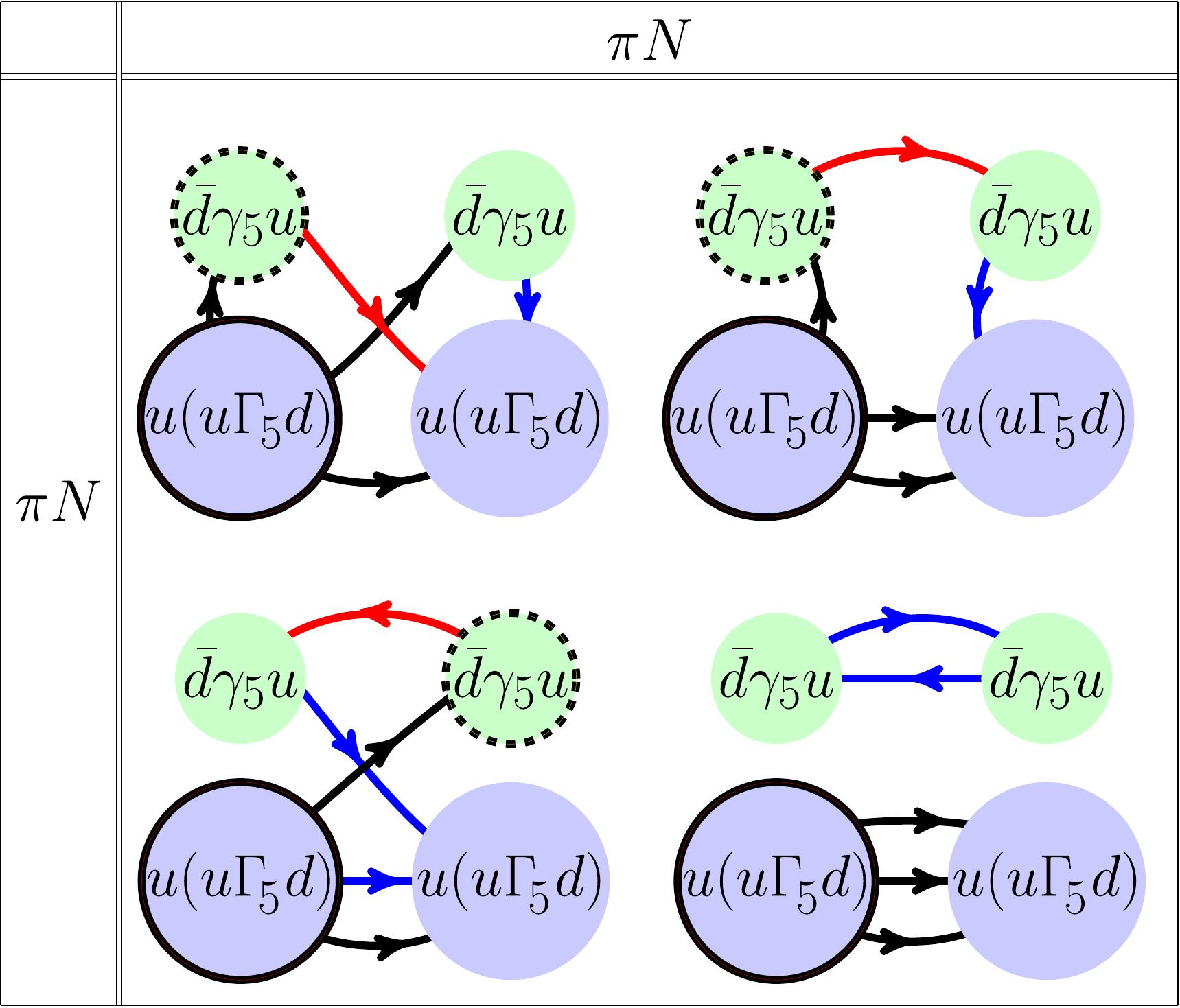}
  \caption{Upper panel: Two-point function contractions involving the $\Delta$ interpolator.
    A gray filling of a circle represents the $\Delta$ interpolator, a green filling represents the $\pi$ interpolator, and a blue filling represents the $N$ interpolator.
    A solid black outline indicates a point source, while a dotted outline represents a sequential source. The black arrow lines represent point-to-all propagators, and the red arrow lines represent sequential propagators. The contractions with the $\pi N$ operator at the sink and the $\Delta$ operator at the source are not computed directly but are obtained from the contraction with the $\Delta$ operator at the sink and the $\pi N$ operator at the source through conjugation. Lower panel: Two-point function contractions for $\pi N - \pi N$. The blue arrow lines represent stochastic propagators, while the other elements are analogous to the upper panel.}
  \label{fig:contract1}
\end{figure}

From the $\Delta/N\pi$ interpolators discussed above, we build two-point correlation matrices for each total momentum $\vec{P}$ and irrep $\Lambda$,
\begin{equation}
  C_{ij}^{\Lambda,\vec{P}}=\langle O_i^{\Lambda,\vec{P}}(t_{snk}) \bar{O}_j^{\Lambda,\vec{P}}(t_{src}) \rangle,
\end{equation}
where the indices $i,j$ now label all the different operators in the same irrep that can vary in internal momentum content, embedding from the multiplicity, or gamma matrices used in the diquarks of Eqs.~(\ref{eq:nucleon_op}) or (\ref{eq:delta_op}).
The Wick contractions are computed following the scheme outlined in Refs.~\cite{Alexandrou:2017mpi,Rendon:2020rtw}.
The correlators with single-hadron interpolators at source and sink are constructed from point-to-all propagators, while the
correlators with a single-hadron interpolator at the sink and a two-hadron $N\pi$ interpolator at the source use in addition a sequential propagator, with sequential inversion
through the pion vertex at source time. The topologies of these diagrams are shown in the top panel of Fig.~\ref{fig:contract1}.
The bottom panel of Fig.~\ref{fig:contract1} shows the topologies for the correlators with $N\pi$ operators at both source and sink.
The diagrams are split into two factors, separated at the source point and by using a stochastic source - propagator pair.
For the latter we use stochastic timeslice sources in the upper two diagrams. In the lower diagrams we employ spin-dilution and the one-end-trick in addition to time dilution.

The quark propagators of all types are Wuppertal-smeared \cite{Gusken:1989ad} at source and sink
with smearing parameters $\alpha_{\mathrm{Wup}} = 3.0$ and $N_{\mathrm{Wup}} = 45$;
these parameters were originally optimized for the nucleon two-point functions in Ref.~\cite{green2014nucleon}. The gauge field deployed in the smearing kernel
is again 2-level HEX-smeared \cite{Hasenfratz:2001hp,Morningstar:2003gk}.

\section{Spectra results}\label{sec_spectra_results}
%
\begin{figure}
  \includegraphics[width=\columnwidth]{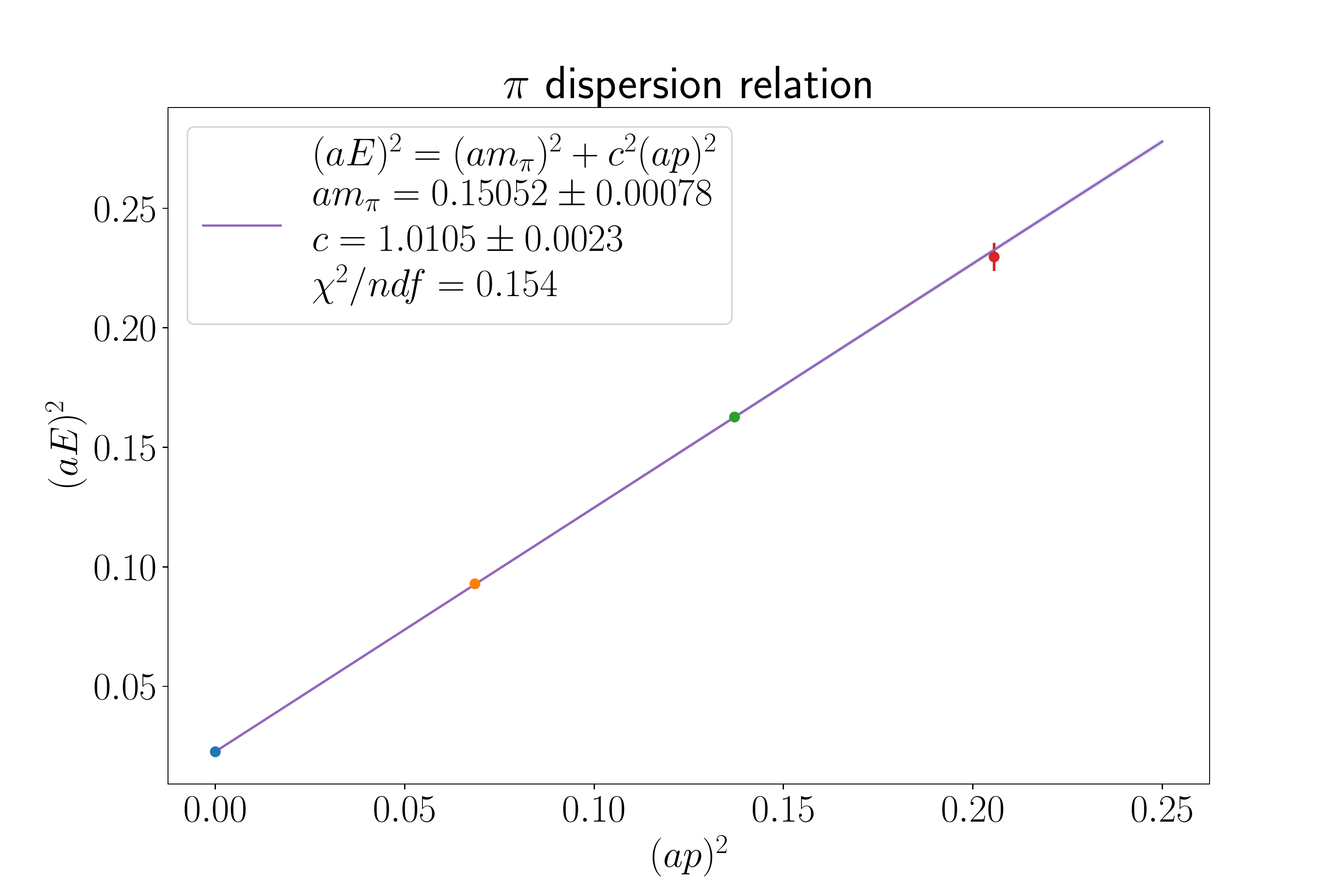}
  \caption{Pion dispersion relation.\label{fig:dispersion_rel_pi}}
\end{figure}
\begin{figure}
  \includegraphics[width=\columnwidth]{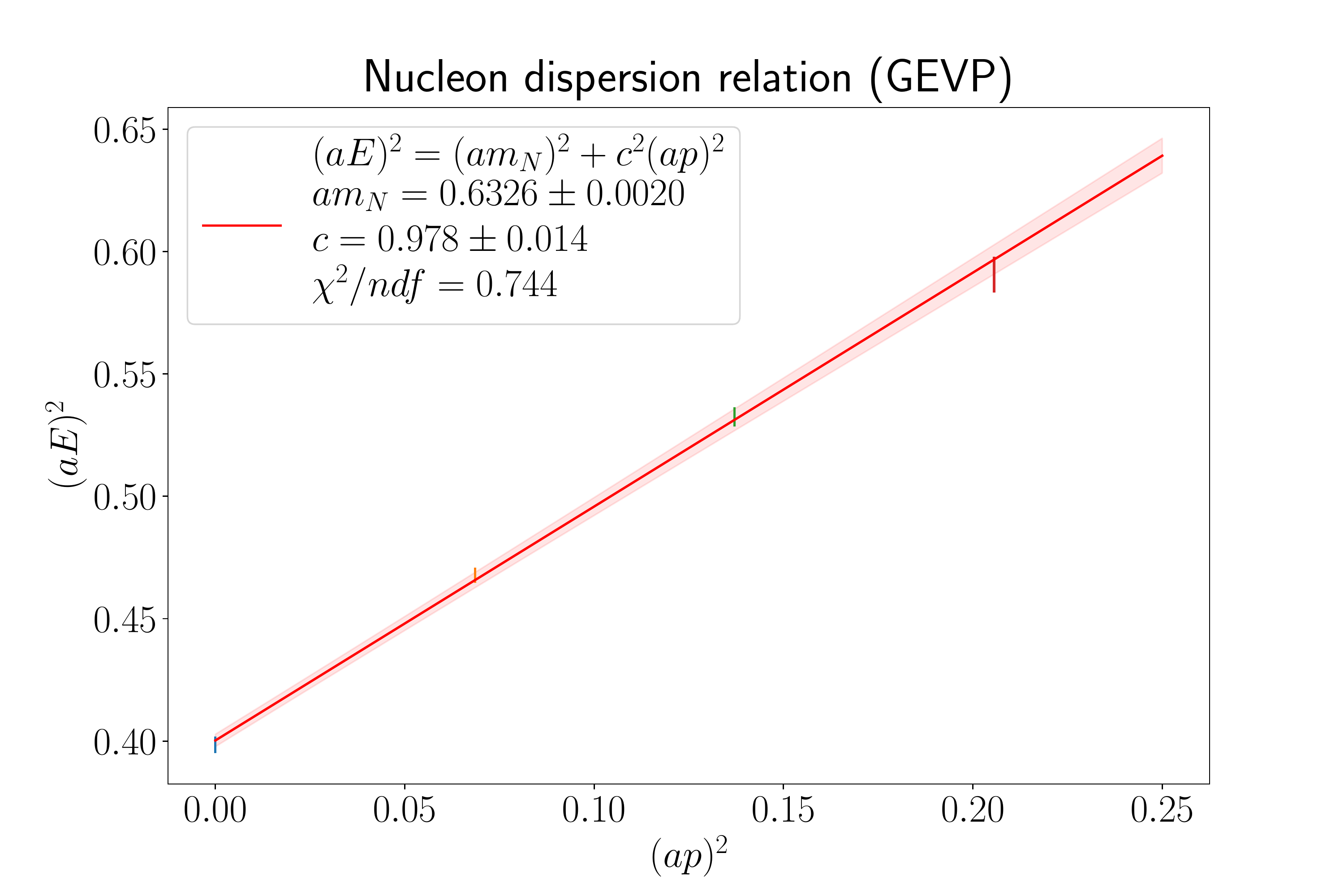}
  \caption{Nucleon dispersion relation from the GEVP analysis.\label{fig:dispersion_rel_N}}
\end{figure}

\setlength{\tabcolsep}{1pt}
\begin{table}
  \begin{tabular}{| c c | c c c |c|}
    \hline
    \rowcolor[gray]{.9}[\tabcolsep]
    $(\frac{L}{2\pi})^2|\vec{P}|^2$ & $\Lambda$ & $n$ & Fit Range & $\frac{\chi^2}{\text{dof}}$ & $a\sqrt{s^{\Lambda,\vec{P}}_n}$ \\
    \hline
    $0$                             & $G_{1u}$  & $1$ & $4-15$    & $1.90$                      & $0.782(4)(3)$                      \\
    $0$                             & $G_{1u}$  & $2$ & $4-15$    & $0.80$                      & $0.978(12)(1)$                     \\
    \hline
    $0$                             & $H_{g}$   & $1$ & $5-15$    & $1.79$                      & $0.829(4)(2)$                      \\
    $0$                             & $H_{g}$   & $2$ & $4-15$    & $0.43$                      & $1.028(6)(4)$                      \\
    \hline
    $1$                             & $G_{1}$   & $1$ & $4-15$    & $1.97$                      & $0.790(5)(4)$                      \\
    $1$                             & $G_{1}$   & $2$ & $5-15$    & $1.14$                      & $0.829(5)(8)$                      \\
    $1$                             & $G_{1}$   & $3$ & $5-15$    & $0.72$                      & $0.914(8)(9)$                      \\
    \hline
    $1$                             & $G_{2}$   & $1$ & $5-15$    & $0.48$                      & $0.827(5)(5)$                      \\
    $1$                             & $G_{2}$   & $2$ & $4-15$    & $0.89$                      & $1.020(7)(18)$                      \\
    \hline
    $2$                             & $(2)G$    & $1$ & $4-15$    & $1.73$                      & $0.795(5)(17)$                      \\
    $2$                             & $(2)G$    & $2$ & $4-15$    & $1.72$                      & $0.826(5)(8)$                      \\
    $2$                             & $(2)G$    & $3$ & $4-15$    & $1.60$                      & $0.839(5)(14)$                      \\
    $2$                             & $(2)G$    & $4$ & $3-15$    & $1.87$                      & $0.917(4)(12)$                      \\
    $2$                             & $(2)G$    & $5$ & $3-15$    & $0.71$                      & $0.939(4)(3)$                      \\
    \hline
    $3$                             & $G$       & $1$ & $3-15$    & $1.32$                      & $0.791(5)(2)$                      \\
    $3$                             & $G$       & $2$ & $3-15$    & $0.68$                      & $0.843(7)(7)$                      \\
    $3$                             & $G$       & $3$ & $3-15$    & $2.01$                      & $0.940(7)(15)$                      \\
    \hline
    $3$                             & $F_{1}$   & $1$ & $4-15$    & $1.46$                      & $0.831(7)(29)$                      \\
    $3$                             & $F_{1}$   & $2$ & $4-15$    & $0.27$                      & $0.960(11)(3)$                     \\
    \hline
    $3$                             & $F_{2}$   & $1$ & $4-15$    & $0.45$                      & $0.839(7)(6)$                      \\
    $3$                             & $F_{2}$   & $2$ & $4-15$    & $0.56$                      & $0.962(6)(7)$                      \\
    \hline
  \end{tabular}
  \caption{Center-of-mass energies in the $\Delta$-$N\pi$ sector from single-exponential fits to the principal correlators, for the different total momenta $\vec{P}$ and irreps $\Lambda$. The first uncertainty is statistical and the second uncertainty is systematic, given by the shift in the fitted energy when increasing $t_{\rm min}$ by one unit.\label{tab:spectrum}}
\end{table}

\begin{figure*}
  \includegraphics[width=\columnwidth]{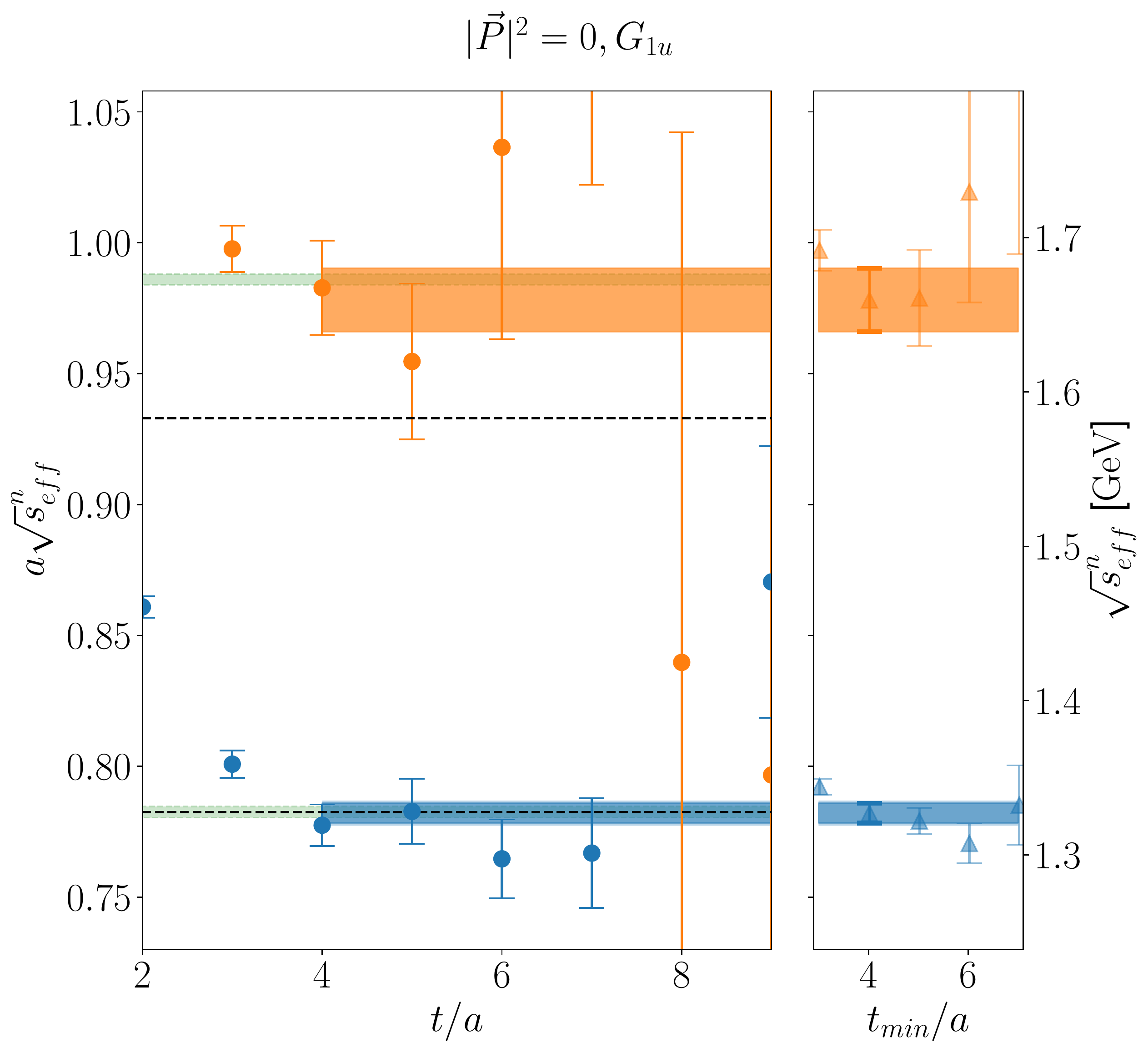}
  \includegraphics[width=\columnwidth]{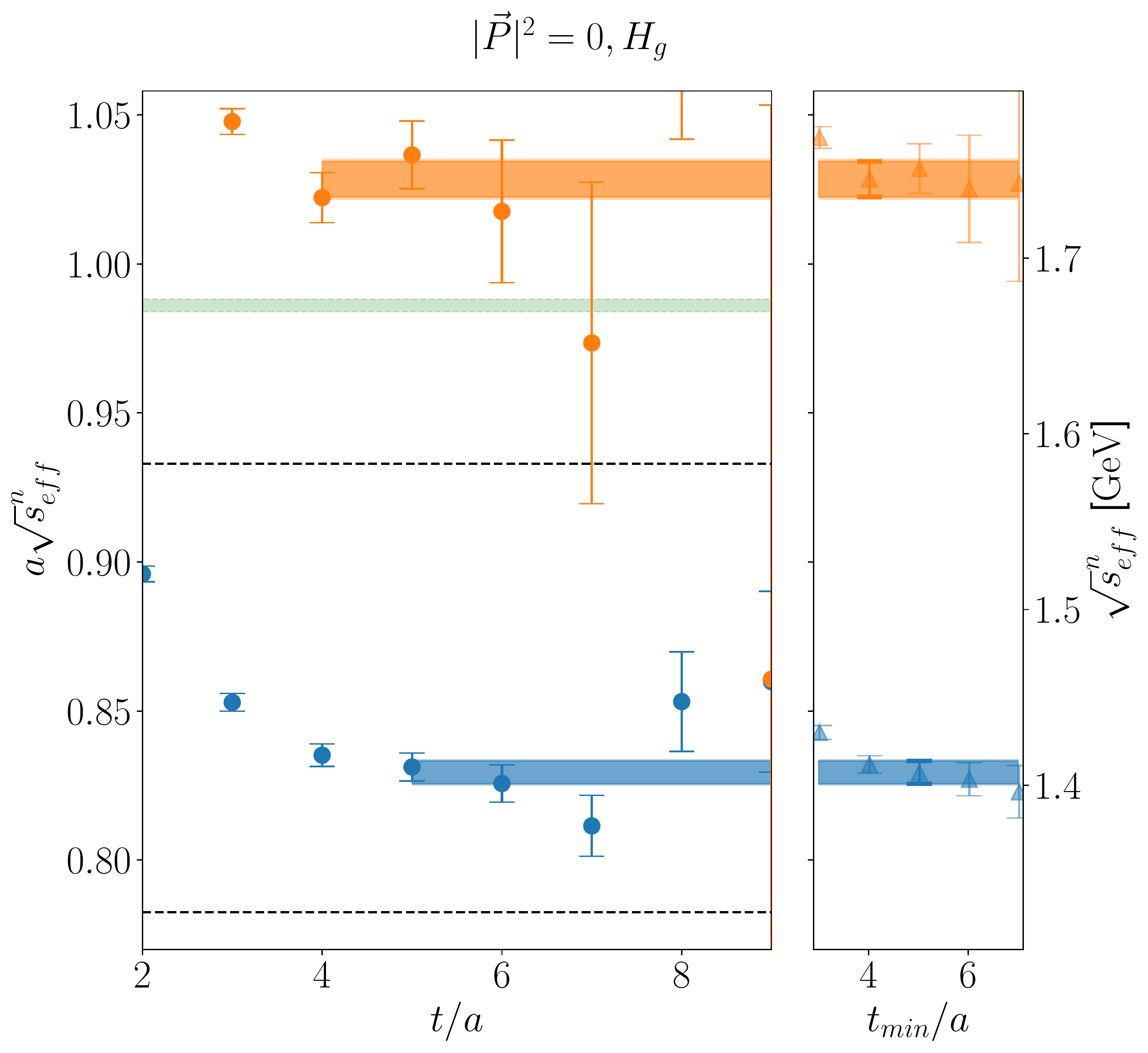}
  \includegraphics[width=\columnwidth]{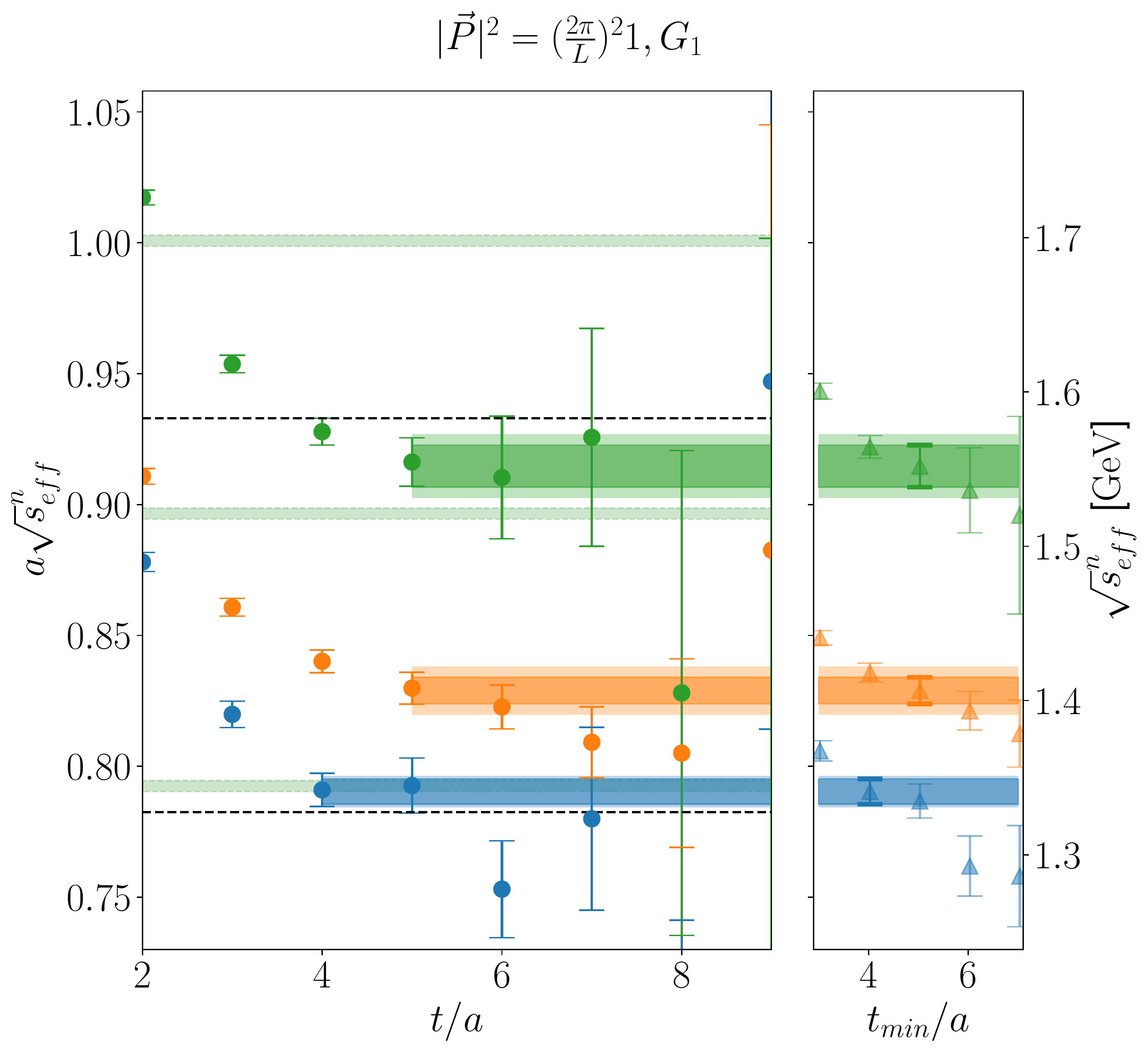}
  \includegraphics[width=\columnwidth]{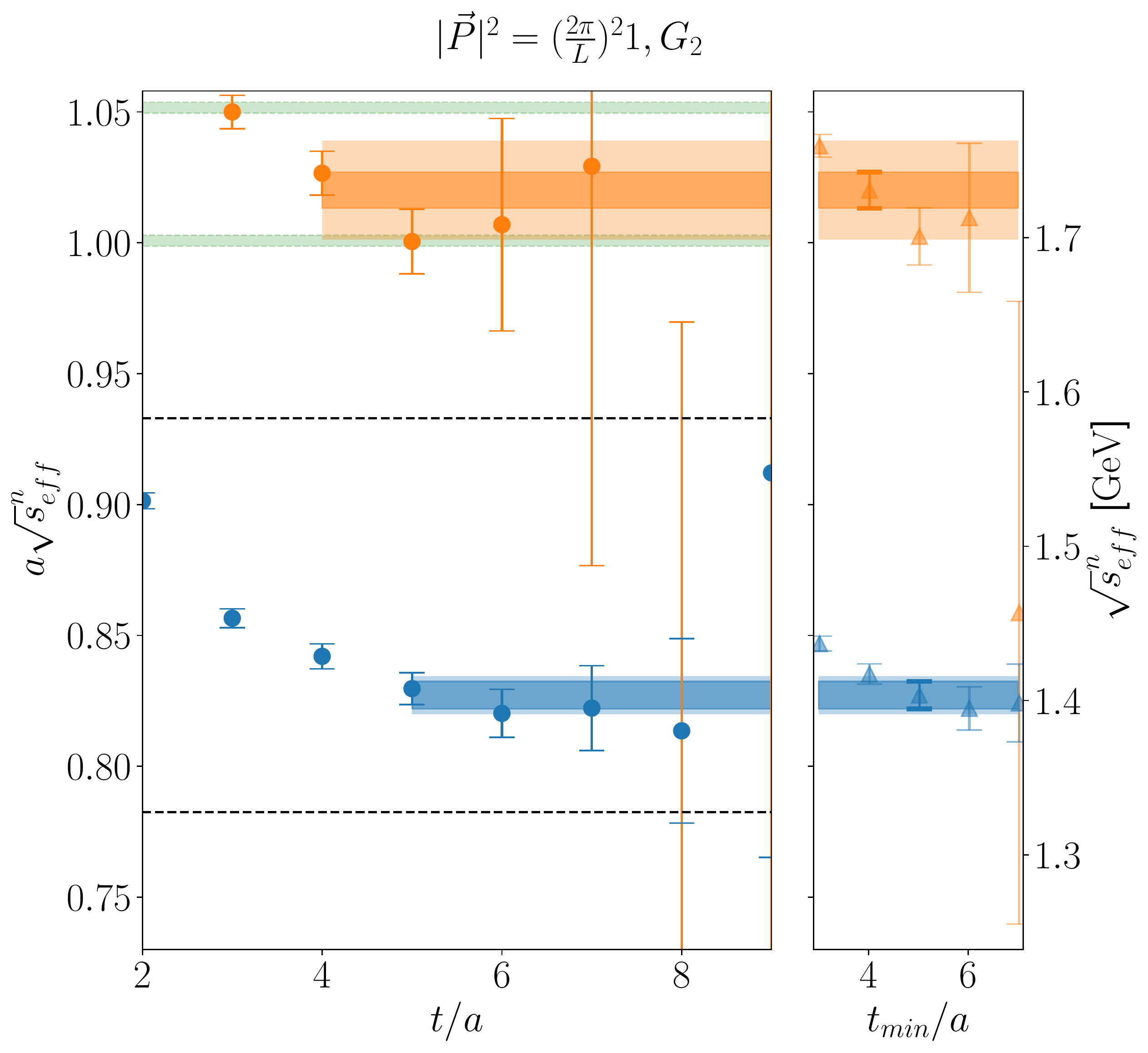}
  \caption{For each irrep, we show the effective energies of the principal correlators as a function of $t/a$ (left), and the energies obtained from single-exponential fits
    to these correlators as a function of $t_{\rm min}/a$ (right). The outer, lighter-shaded bands include an estimate of the systematic uncertainty associated with the choice of fit range, calculated from the change in the fitted energy when increasing $t_{\rm min}/a$ by $+1$. All energies shown here are converted to the center-of-mass frame. Black dashed lines represent the $N\pi$ and $N\pi\pi$ thresholds. Non-interacting $N\pi$ energy levels are shown as green lines. \label{fig:meff_plot} }
\end{figure*}

\begin{figure*}
  \includegraphics[width=\columnwidth]{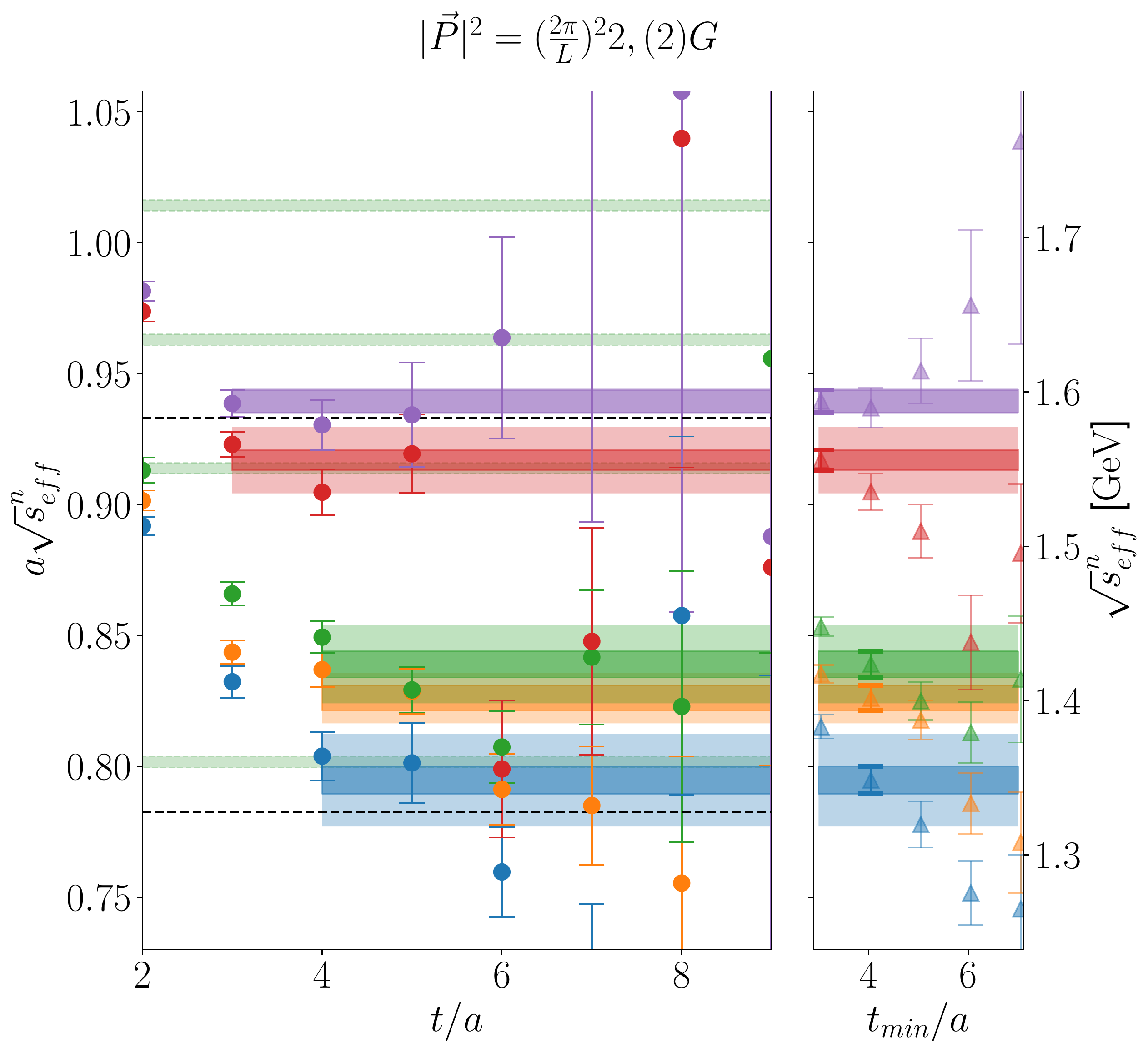}
  \includegraphics[width=\columnwidth]{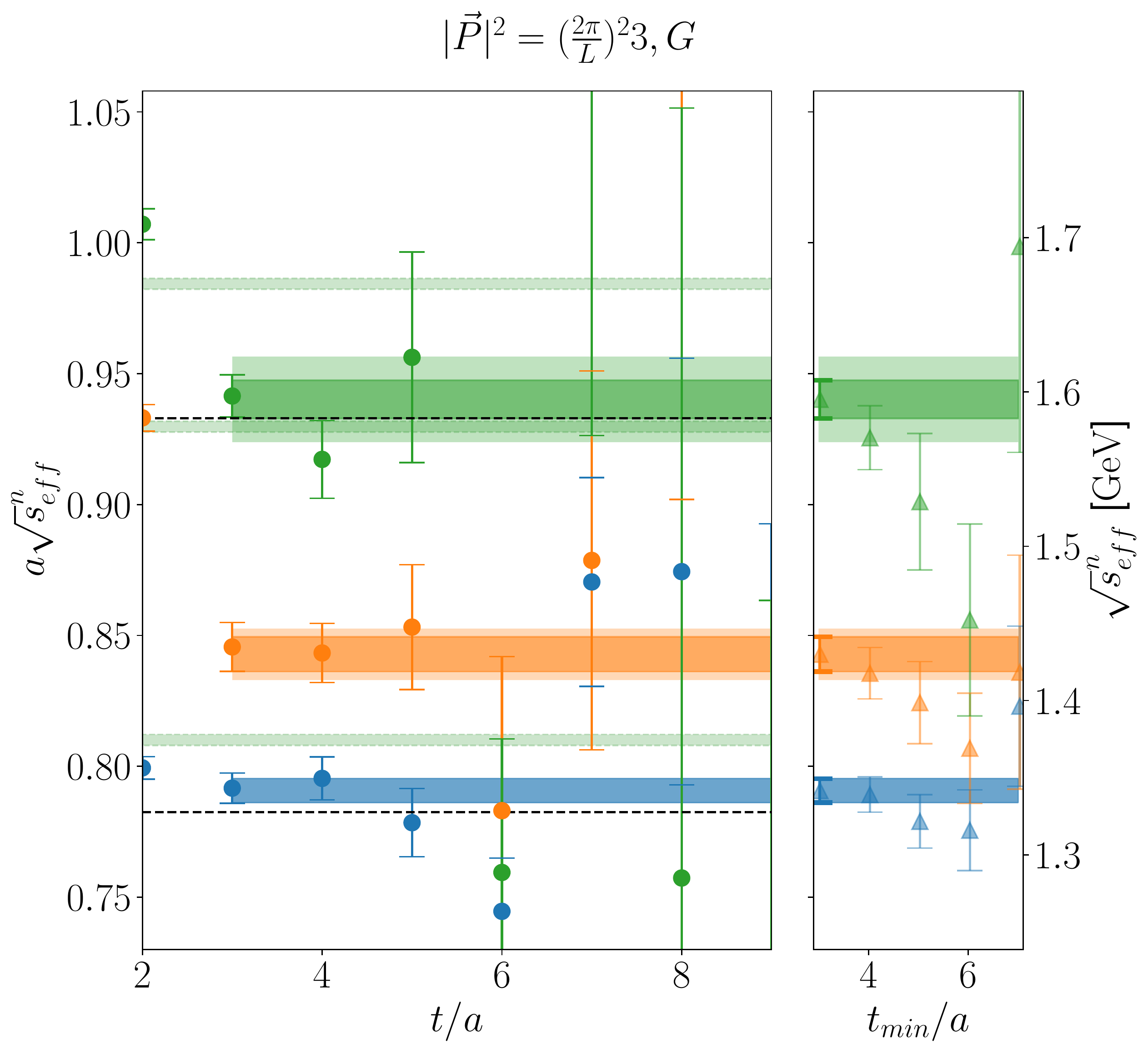}
  \includegraphics[width=\columnwidth]{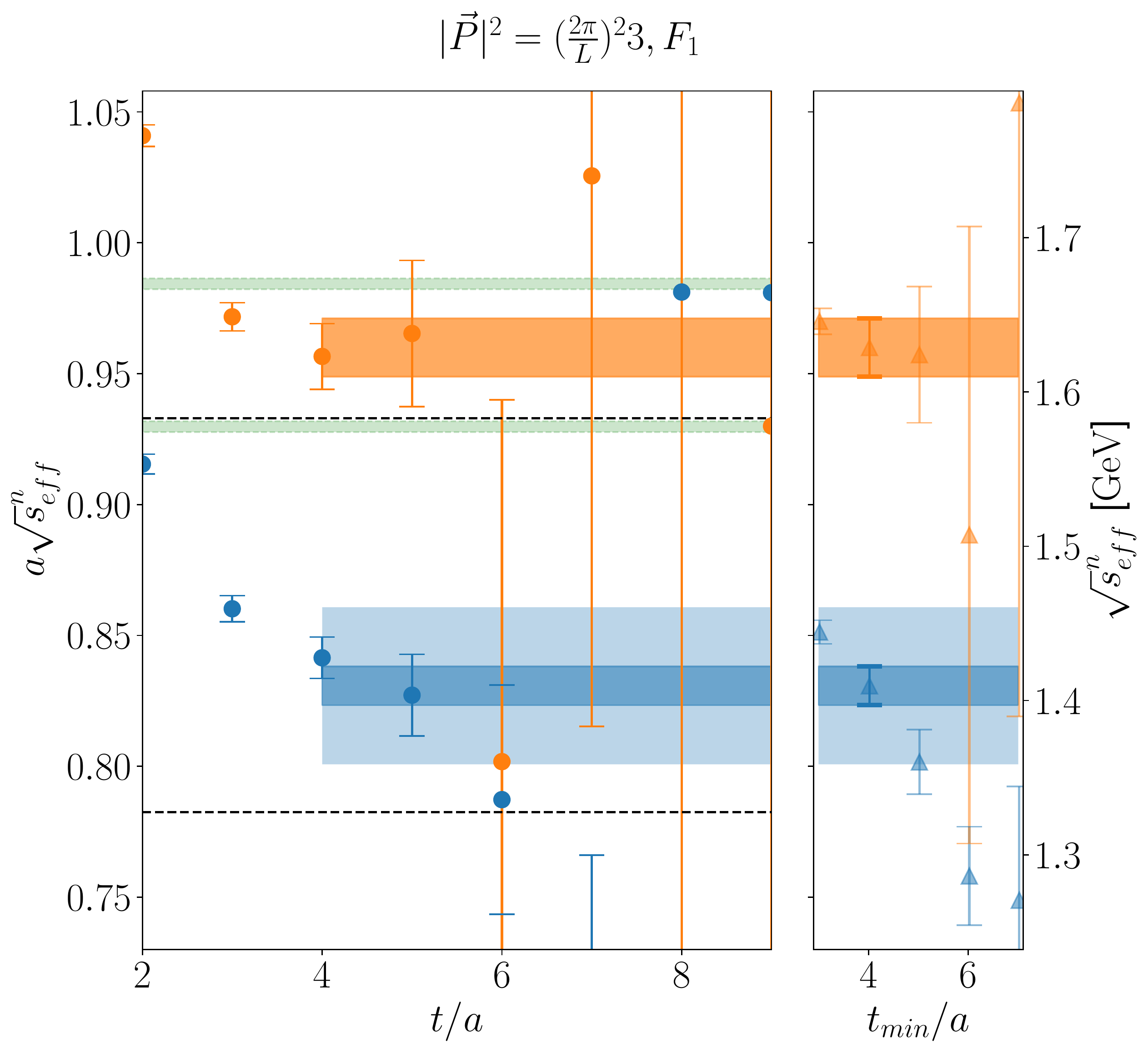}
  \includegraphics[width=\columnwidth]{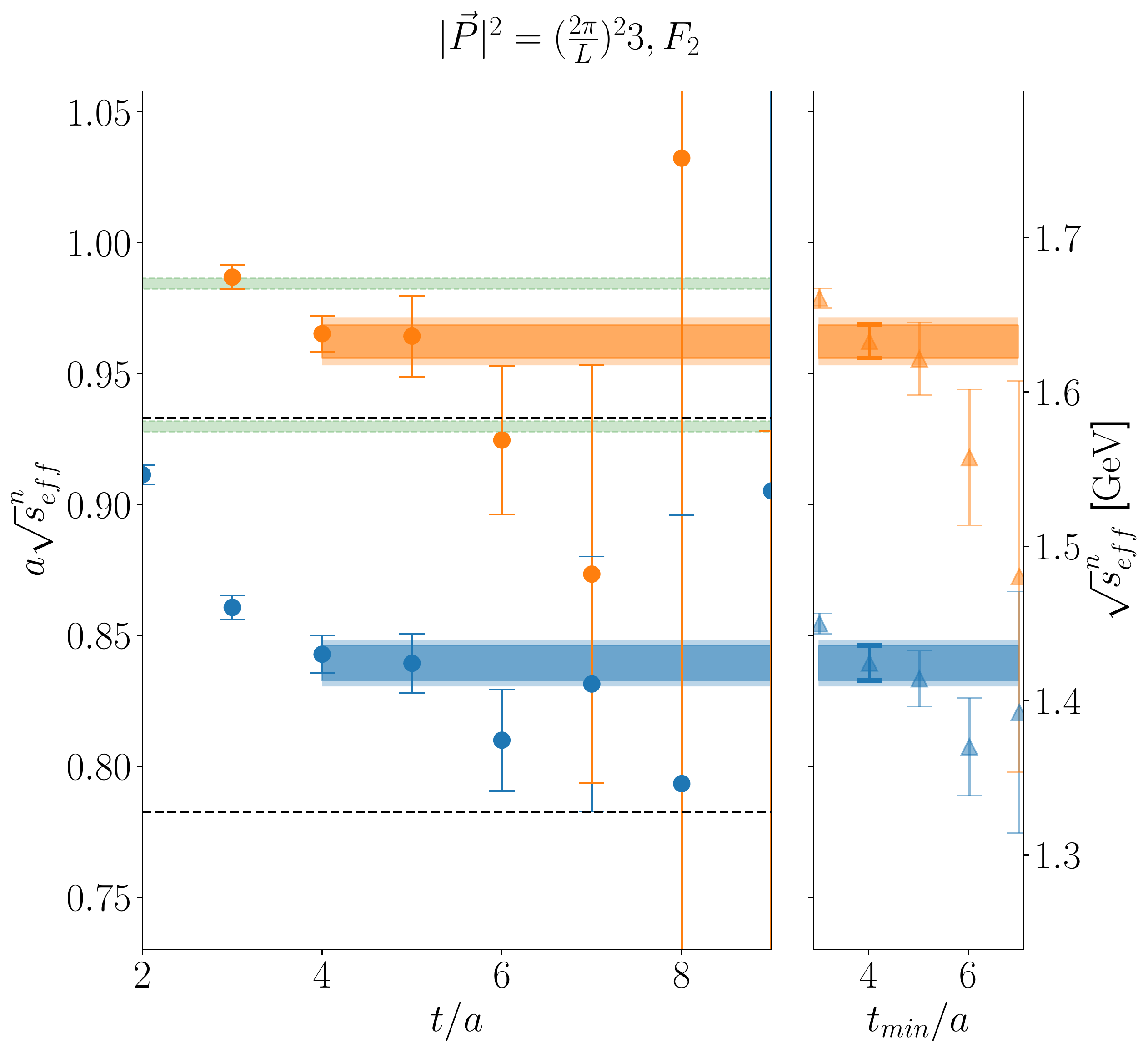}
  \caption{Like Fig.~\ref{fig:meff_plot}, but with irreps $(2)G,G,F_1,F_2$. \label{fig:meff_plot_2}}
\end{figure*}

\begin{figure*}[t]
  \includegraphics[width=0.8\linewidth]{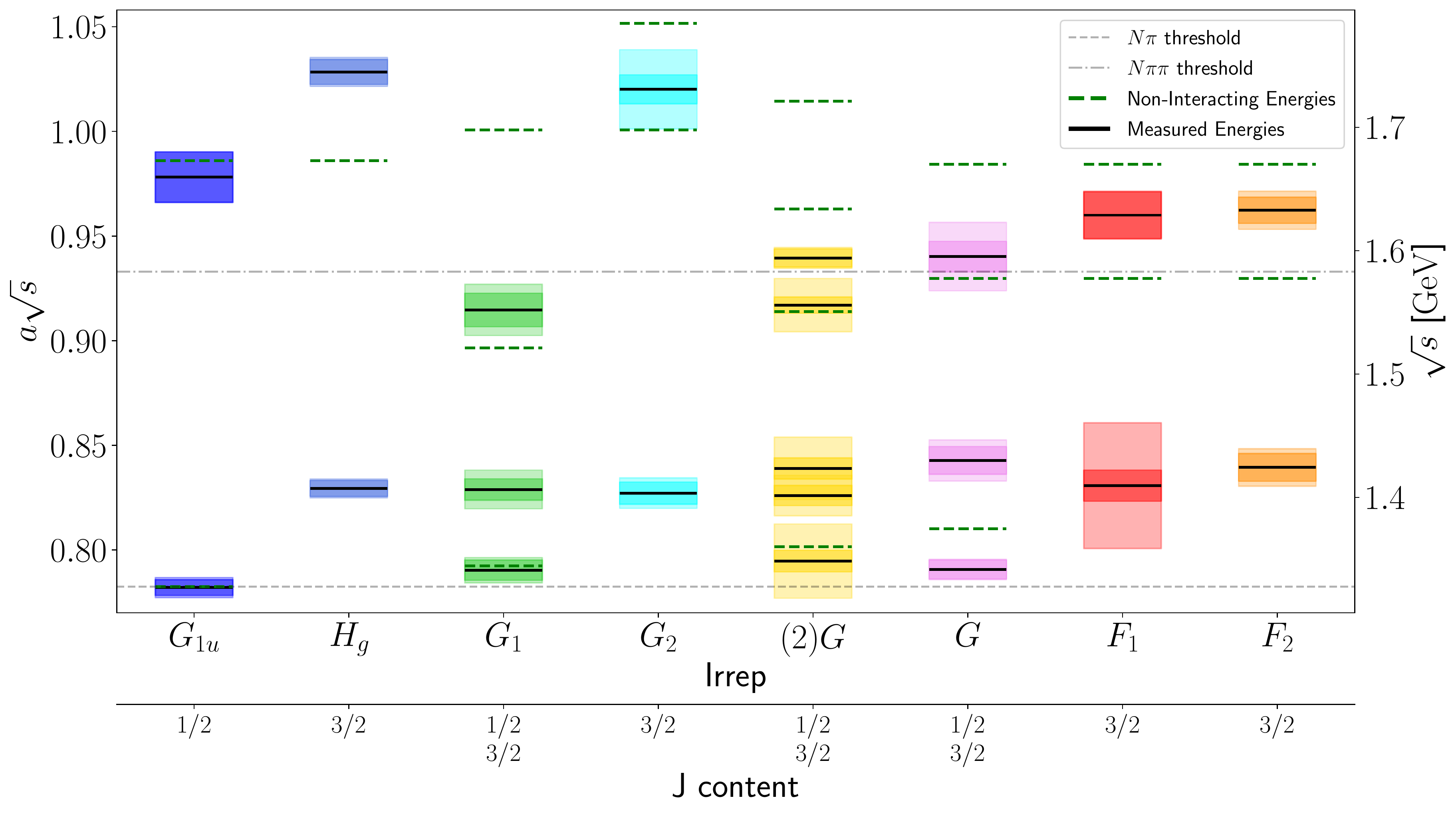}
  \caption{
    Energy levels extracted in each irrep, with $J\leq3/2$ content listed. The inner bands indicate the statistical and scale-setting uncertainties. The outer, lighter-shaded bands include an estimate of the systematic uncertainty associated with the choice of fit range, calculated from the change in the fitted energy when increasing $t_{\rm min}/a$ by $+1$.\label{fig:all_irrep_spectra}
  }
\end{figure*}

The masses of the pion and nucleon are used as input parameters in the L{\"u}scher method.
We extract them from fits of their dispersion relations, shown in Figs.~\ref{fig:dispersion_rel_pi} and \ref{fig:dispersion_rel_N}, giving
\begin{eqnarray}
  a m_\pi  &=&  0.15052(78), \\
  a m_N &=&  0.6326(20).
\end{eqnarray}
The energies are obtained from single-state fits of the two-point functions projected to different momenta (using a $\cosh$ for the pion and a single exponential for the nucleon).

For the $\Delta$-$N\pi$ system, to extract the energy levels $E_n^{\Lambda,\vec{P}}$ (where $n$ now counts the finite-volume energy levels for a given $\Lambda,\vec{P}$) from the correlation matrices $C_{ij}^{\Lambda,\vec{P}}$ we use the generalized eigenvalue problem (GEVP) \cite{michael1985adjoint,luscher1990calculate,blossier2009generalized,orginos2015improved}
\begin{equation}
  C_{ij}^{\Lambda,\vec{P}}(t)u_j^n(t)=\lambda^n(t,t_0)C_{ij}^{\Lambda,\vec{P}}u_j^n(t),
\end{equation}
where  $u^n_j$ are the right generalized eigenvectors.
In the plateau regions the energies are obtained from fits to the principal correlators $\lambda^n(t,t_0)$ with single exponentials as
\begin{equation}
  \lambda^n (t,t_0) \sim e^{-E_n^{\Lambda,\vec{P}} (t-t_0)}.
\end{equation}
Here, $t_0$ is a reference timeslice that does not strongly affect the large-$t$ behavior; we set $t_0/a=2$.

Additionally, for the projected multihadron operators $(N\pi)^{\Lambda,r}(\vec{P})$ we implement an optimized interpolator of the nucleon \cite{Dudek:2012gj}
\begin{equation}
  \mathcal{N}(\vec{p}_1)=\sum_i u_{(N)i}^{1}(t)N_{i}(\vec{p}_1), \label{eq:nucleonoptimized}
\end{equation}
where $i$ labels the two types of nucleon operators in Eq.~(\ref{eq:nucleon_op}) and $u_{(N)i}$ are the generalized eigenvectors (for $t/a=4$) from a single-nucleon GEVP analysis.
The optimized nucleon interpolator has improved overlap with the single-nucleon ground state with momentum $\vec{p}_1$ \cite{alexandrou2018pi}.

For the coupled $\Delta$-$N\pi$ system we build for each irrep $\Lambda$
a correlation matrix $C_{ij}^{\Lambda,\vec{P}}$ from the projected $\Delta$ and optimized $\mathcal{N}\pi$ operators in Table \ref{tab:all_irreps}.
The multiplicities of operators give rise to a fairly large basis for each correlation matrix (the dimensions for the full bases correspond to the sums of numbers of operators for each irrep listed in Table \ref{tab:all_irreps}). Through singular value decomposition of $CC^{\dag}$ or $C^{\dag}C$ we can infer which operators contribute to the largest singular values, allowing us to explore sub-bases of the full list of operators that can lead to reduced noise of the principal correlators while maintaining the complete spectra.

Baryons are known to have a narrow plateau region (the "golden window" \cite{beane2009high})
where the higher states contribution get small enough to enable a single exponential fit to describe maximally
a single level before the rapid decay of signal-to-noise ratio at larger $t$ \cite{leinweber2005baryon,wagman2017statistics}.
In the left subplot for each irrep in Figs.~\ref{fig:meff_plot} and \ref{fig:meff_plot_2}, we show the
the effective masses of the principal correlators,
\begin{equation}
  aE_{eff}^n (t)= \ln \frac{\lambda_n(t,t_0)}{\lambda_n(t+a,t_0)},
\end{equation}
converted to the center-of-mass frame using
\begin{equation}
  \sqrt{s_n}^{\Lambda,\vec{P}}= \sqrt{(E_n^{\Lambda,\vec{P}})^2-(\vec{P})^2}.
\end{equation}
The center-of-mass energies are also related to the scattering momenta through
\begin{equation}
  \sqrt{s_n}^{\Lambda,\vec{P}}= \sqrt{(k_n^{\Lambda,\vec{P}})^2+m_{\pi}^2}+\sqrt{(k_n^{\Lambda,\vec{P}})^2+m_N^2}.\label{eq:scatteringmom}
\end{equation}
Our main results are obtained from single-exponential fits to the principal correlators and are listed in Table \ref{tab:spectrum}.
The fit ranges are chosen after a stability analysis. The upper limit of the fit range, once chosen large enough, is found to have a small impact on the fit itself; thus, we fix it to $t_{max}/a=15$ for all levels. On the other hand, the lower limit is varied within a reasonable range until a plateau region is identified. This is illustrated in the right subplot for each irrep in Figs.~\ref{fig:meff_plot} and \ref{fig:meff_plot_2}. In addition, we
estimate a systematic uncertainty for each energy level as the shift in the fitted energy when increasing $t_{\rm min}/a$ by +1. These uncertainties have been added in quadrature in the lighter-shaded outer bands shown in Figs.~\ref{fig:meff_plot} and \ref{fig:meff_plot_2}, and will also be propagated to the scattering amplitudes in Sec.~\ref{sec_result}.

To further test the stability, we also attempted two-exponential fits using the form
\begin{equation}
  \lambda^n (t,t_0) \sim (1-B)e^{-E_n^{\Lambda,\vec{P}} (t-t_0)} + B e^{-E^{\prime \Lambda,\vec{P}}_n (t-t_0)},
\end{equation}
where $E^{\prime \Lambda,\vec{P}}_n$ would be a high-lying energy level not covered by the GEVP analysis.
These fits give consistent results for $E_n^{\Lambda,\vec{P}}$, but the results for the
parameters $B$ and $E^{\prime \Lambda,\vec{P}}_n$ are rather unstable under variations of $t_{\rm min}$ at our level of correlator precision.

It can be seen in the plots that energy levels that overlap strongly with the $N\pi$ states shift away from the resonance region, as expected.
For the irrep $(2)G$ in $|\vec{P}|^2=(2\pi/L)^2 2$, the situation is more complicated and a higher number of energy states appear in the region of interest.
This situation originates from having only a single irrep for the Little Group $C_{2v}^D$, resulting in a maximal mixing of angular momenta.

A summary of all extracted energy levels is shown in Fig.~\ref{fig:all_irrep_spectra}.

\section{L{\"u}scher quantization conditions}\label{sec_luscher_analysis}

The L{\"u}scher quantization condition connects the finite-volume energy spectra affected by the interactions and the infinite-volume scattering amplitudes; resonances correspond to poles in the infinite-volume scattering amplitudes at complex $\sqrt{s}$ and in principle affect the entire spectrum. For elastic 2-body scattering of nonzero-spin particles, the quantization condition
can be written as \cite{gockeler2012scattering}
\begin{equation}\label{eq:QC}
  \det ( \mathcal{M}^{\vec{P}}_{Jl\mu,J^\prime l^\prime \mu^\prime} - \delta_{JJ^\prime} \delta_{ll^\prime} \delta_{\mu\mu^\prime} \cot \delta_{Jl}) = 0,
\end{equation}
where $\delta_{Jl}$ is the infinite-volume scattering phase shift for total angular momentum $J$ and orbital angular momentum $l$, and $\mu,\mu^\prime=-J,...,J$. Both the
scattering phase shift and the matrix ${\cal M}^{\vec{P}}_{Jl\mu,J^\prime l^\prime \mu^\prime}$ are functions of the scattering momentum, and the solutions of the quantization condition for the scattering momentum give the finite-volume energy levels through Eq.~(\ref{eq:scatteringmom}). The matrix ${\cal M}^{\vec{P}}_{Jl\mu,J^\prime l^\prime \mu^\prime}$ encodes the geometry of the finite box and is a generalization for particles with spins $\sigma,\sigma^\prime$ of the spinless counterpart via
\begin{align} \label{eq:mat_Jlmu}
  \mathcal{M}^{\vec{P}}_{\substack{Jl\mu,                                                                          \\ J^\prime l^\prime \mu^\prime}} =& \sum_{\substack{m,\sigma, \\ m^\prime, \sigma^\prime}}\braket{lm,\frac{1}{2} \sigma | J \mu}
  \braket{l^\prime m^\prime,\frac{1}{2} \sigma^\prime | J^\prime \mu^\prime} \mathcal{M}^{\vec{P}}_{\substack{l m, \\ l^\prime m^\prime}},
\end{align}
where $\mathcal{M}^{\vec{P}}_{l m,l^\prime m^\prime}$ (for a cubix box with periodic boundary conditions) is given by \cite{gockeler2012scattering}
\begin{align}
  \mathcal{M}_{lm,l^\prime m^\prime}^{\vec{P}}(q^2)= & \frac{(-1)^l \gamma^{-1}}{\pi^{3/2}} \sum_{j=\mid l-l^\prime \mid}^{l+l^\prime} \sum_{s=-j}^{j} \frac{i^j}{q^{j+1}} \nonumber \\
                                                     & \times Z^{\vec{P}}_{js}(1,q^2)^\ast C_{lm,js,l^\prime m^\prime},
\end{align}
where $q=\frac{k L}{2\pi}$ with $k$ the scattering momentum and $L$ the side length of the box.
\begin{table*}
  \begin{tabular}{|c c | c | c|}
    \hline
    \rowcolor[gray]{.9}[\tabcolsep]
    $\frac{L}{2\pi}\vec{P}$    & Group $LG$                  & Irrep $\Lambda$ & Quantization condition                                                                                         \\
    \hline
    \multirow{2}{*}{$(0,0,0)$} & \multirow{2}{*}{$O_{h}^D$}  & $G_{1u}$        & $-w_{00}+\cot\delta_{\frac{1}{2},0}=0$                                                                         \\
    \cline{3-4}
                               &                             & $H_{g}$         & $-w_{00}+\cot\delta_{\frac{3}{2},1}=0$                                                                         \\
    \hline
    \multirow{2}{*}{$(0,0,1)$} & \multirow{2}{*}{$C_{4v}^D$} & $G_{1}$         & $-2w_{10}^2+(w_{00} - \cot\delta_{\frac{1}{2},0})(w_{00} + w_{20} -\cot\delta_{\frac{3}{2},1})=0 $             \\
    \cline{3-4}
                               &                             & $G_{2}$         & $-w_{00}+w_{20}+\cot\delta_{\frac{3}{2},1}=0 $                                                                 \\
    \hline
    $(1,1,0)$                  & $C_{2v}^D$                  & $(2)G$          & \makecell{$-(w_{00}-\cot\delta_{\frac{1}{2},0})(-w_{20}^2 + 2w_{22}^2 +(w_{00}-\cot\delta_{\frac{3}{2},1})^2)$ \\ $-4 \text{Re}(w_{11})^2 (2w_{00}+w_{20}-i \sqrt{6}w_{22}-2\cot\delta_{\frac{3}{2},1}) =0$ } \\
    \hline
    \multirow{2}{*}{$(1,1,1)$} & \multirow{2}{*}{$C_{3v}^D$} & $G$             & $-6w_{10}^2+(w_{00} - \cot\delta_{\frac{1}{2},0})(w_{00} -i\sqrt{6}w_{22} -\cot\delta_{\frac{3}{2},1})=0$      \\
    \cline{3-4}
                               &                             & $F_1, F_2$      & $-w_{00}-i\sqrt{6}w_{22}+\cot \delta_{\frac{3}{2},1}=0$                                                        \\
    \hline
  \end{tabular}
  \caption{Finite-volume quantization conditions for all irreps in terms of phase shifts $\delta_{J,l}$ and functions $w_{lm}$. \label{tab:qc}}
\end{table*}
Here $Z^{\vec{P}}_{js}(1,q^2) $ is the generalized zeta function, $\gamma=E^{\vec{P}}/\sqrt{s} $ is the Lorentz boost factor
and the coefficient $C_{lm,js,l^\prime m^\prime}$ expressed in terms of Wigner 3$j$-symbols read
\begin{gather}
  C_{lm,js,l^\prime m^\prime}=(-1)^{m^\prime} i^{l-j-l^\prime} \sqrt{(2l+1)(2j+1)(2l^\prime+1)}  \nonumber \\
  \times
  \begin{pmatrix}
    l & j & l^\prime  \\
    m & s & -m^\prime
  \end{pmatrix}
  \begin{pmatrix}
    l & j & l^\prime \\
    0 & 0 & 0
  \end{pmatrix}.
\end{gather}
To simplify notation it is common practice to define the functions
\begin{equation} \label{wlm}
  w_{lm}= w_{lm}^{\vec{P}}(q,L) \equiv \frac{Z_{lm}^{\vec{P}}(1;q^2)}{\gamma \pi^{3/2} \sqrt{2l+1}  q^{l+1}}.
\end{equation}
The elements of the matrices $\mathcal{M}_{J l \mu, J' l' \mu'}^{\vec{P}}$ for all choices of $\vec{P}$ considered in this work are listed in Appendix \ref{app:matrixM}.

Furthermore, it is possible to extract quantization conditions for each irrep $\Lambda$ via a change of basis of Eq.~(\ref{eq:QC}). The basis vector of the irrep $\Lambda$ can be written as \cite{gockeler2012scattering,romero2018vector}
\begin{equation}
  \ket{\Lambda r J l n}=\sum_{\mu} c_{Jl\mu}^{\Lambda r n} \ket{Jl\mu},
\end{equation}
where the coefficients $c_{Jl\mu}^{\Lambda r n}$ for $l\leq 2$ can be found in Refs.~\cite{gockeler2012scattering,bernard2008resonance}, and the parity eigenstate vectors $\ket{Jl\mu}$ are given by

\begin{equation}
  \ket{Jl\mu}=\sum_{m,\sigma} \ket{lm, \frac{1}{2} \sigma} \braket{lm, \frac{1}{2} \sigma|J\mu}.
\end{equation}
One can then make a change of basis for which the matrix elements of $\mathcal{M}$ are given by
\begin{align} \label{eq:changeofbasis}
  \bra{\Lambda r J l n} \mathcal{M} \ket{\Lambda^\prime r^\prime J^\prime l^\prime n^\prime} & = \sum_{\mu \mu^\prime} c_{J l \mu}^{\Lambda r n } c_{J^\prime l^\prime \mu^\prime}^{\Lambda^\prime r^\prime n^\prime} \mathcal{M}_{Jln,J^\prime l^\prime n^\prime} \nonumber \\
                                                                                             & = \delta_{\Lambda \Lambda^\prime} \delta_{r r^\prime }\mathcal{M}_{Jln,J^\prime l^\prime n^\prime},
\end{align}
where it is found, from Schur's lemma, that the matrix $\mathcal{M}$ is partially diagonalized in irrep $\Lambda$ and row $r$.
However, the matrix is not diagonal in $n$, which labels the multiple embeddings of the irreps. In our case only the irrep $(2)G$ of the group $C_{2v}^D $ has multiple embeddings with multiplicity $m_G=2$ (See Table \ref{tab:plan}).

In principle, there are infinitely many values of total angular momentum $J$ and therefore also infinitely many partial waves $l$ in each irrep, but, as the higher waves have an increasingly smaller contribution, we consider only the dominant partial waves. In particular, we assume the contributions from partial waves in $J>3/2$ to be negligible and exclude them from the analysis.
For the  $N$-$\pi$ system, $J=3/2$ includes both the $P$-wave $(l=1)$ and the $D$-wave $(l=2)$, with the former being the dominant contribution.
Several irreps mix $J=3/2$ with $J=1/2$, and the latter includes $l=0,1$.


Among the partial wave amplitudes with $J = 1/2$, the $P$-wave ($l = 1$) is expected
to be suppressed relative to the $S$-wave ($l = 0$). At our level of precision,
we find the latter, i.e. $S_{31}$, already to be consistent with zero. Given the
additional suppression of $P_{31}$ relative to $S_{31}$, we decided to not include $P_{31}$ in our present analysis, and this is left for future work.

In addition to the resonant phase shift $P_{33}$ $(J=3/2, l=1)$ for isospin $I=3/2$ we then have only the $S_{31}$ $(J=1/2,l=0)$ , for which the closest resonance would be the distant $\Delta(1620)$.
In order to better constrain the $S_{31}$ $(J=1/2,l=0)$ contribution, we also include the irrep $G_{1u}$, which is the only irrep we can access that contains only spin $J=1/2$ and $l=0$ (up to contributions from $l\geq2$), ensured by the negative parity (\textit{ungerade}). As can be seen in Table \ref{tab:all_irreps}, the interpolating operators in the $G_{1u}$ irrep are exclusively $N$-$\pi$ two-hadron operators, consistent with the expectation that the $S_{31}$ phase shift is nonresonant at low energy.
The quantization conditions for all irreps, expressed in terms of the two phase shifts $\delta_{3/2,1},\delta_{1/2,0}$ and the functions $w_{lm}$, are listed in Table \ref{tab:qc}.

\section{Results for the scattering amplitudes}\label{sec_result}

%
\begin{table*}
  \begin{tabular}{| c|c |c |c | c | c | c |}
    \rowcolor[gray]{.9}[\tabcolsep]
    \hline
    Label                                        & Fit to ($J,l$)         & Irreps $\Lambda$                      & $\sqrt{s}$ points & Breit-Wigner parameters             & ERE parameters        & $\chi^2/{\rm dof}$ \\
    \hline
    \textbf{S}                                   & $(1/2,0)$              & $G_{1u}$                              & 2                 & -                                   & $a_0/a=0.51 \pm 0.96$ & $0.16$             \\
    \hline
    \textbf{P}                                   & $(3/2,1)$              & $H_{g},G_{2},F_{1},F_{2}$             & 8                 & \makecell{$g_{\rm BW}=13.36\pm0.80$                                              \\$am_{\rm BW}=0.8158 \pm 0.0031$ \\
    corr$( a m_{\rm BW}, g_{\rm BW} ) = -0.279$} & -                      & $1.35$                                                                                                                                       \\
    \hline
    \textbf{G(a)}                                & $(1/2,0),(3/2,1)$      & \makecell{$G_{1u},H_{g},G_{1},G_{2},$                                                                                                        \\$(2)G,G,F_{1},F_{2}$} & 21 & \makecell{$g_{\rm BW}=13.62\pm0.50 $\\$am_{\rm BW}=0.8136 \pm 0.0029 $ \\
    corr$( a m_{\rm BW}, g_{\rm BW} ) = -0.375$} & $a_0/a=0.38 \pm 0.44 $ & $0.85$                                                                                                                                       \\
      \hline
    \textbf{G(a+1)}                              & $(1/2,0),(3/2,1)$      & \makecell{$G_{1u},H_{g},G_{1},G_{2},$                                                                                                        \\$(2)G,G,F_{1},F_{2}$} & 21 & \makecell{$g_{\rm BW}=14.05\pm0.83$\\$am_{\rm BW}=0.8088 \pm 0.0043$ \\
    corr$( a m_{\rm BW}, g_{\rm BW} ) = -0.442$} & $a_0/a=0.46\pm 0.80$   & $0.99$                                                                                                                                       \\
    \hline
    \textbf{G(b)}                                & $(1/2,0),(3/2,1)$      & \makecell{$G_{1u},H_{g},G_{1},G_{2},$                                                                                                        \\$(2)G,G,F_{1},F_{2}$} & 21 & \makecell{$g_{\rm BW}=13.54\pm0.59$\\$am_{\rm BW}=0.8161 \pm 0.0030$ \\
    corr$( a m_{\rm BW}, g_{\rm BW} ) =-0.324$}  & $a_0/a=0.34 \pm 0.57$  & $1.56$                                                                                                                                       \\
      \hline
    \textbf{G(c)}                                & $(1/2,0),(3/2,1)$      & \makecell{$G_{1u},H_{g},G_{1},G_{2},$                                                                                                        \\$G,F_{1},F_{2}$} & 15 & \makecell{$g_{\rm BW}=13.67\pm0.57$\\$am_{\rm BW}=0.8146 \pm 0.0030$ \\
    corr$( a m_{\rm BW}, g_{\rm BW} ) = -0.372$} & $a_0/a=0.68 \pm 0.49$  & $0.99$                                                                                                                                       \\
    \hline
    \textbf{G(d)}                                & $(1/2,0),(3/2,1)$      & \makecell{$H_{g},G_{1},G_{2},$                                                                                                               \\$(2)G,G,F_{1},F_{2}$} & 19 & \makecell{$g_{\rm BW}=13.65\pm0.52$\\$am_{\rm BW}=0.8137 \pm 0.0029$ \\
    corr$( a m_{\rm BW}, g_{\rm BW} ) = -0.360$} & $a_0/a=0.52\pm 0.85$   & $0.93$                                                                                                                                       \\
    \hline
  \end{tabular}
  \caption{Fit results for the scattering parameters, using different combinations of energy levels as explained in the main text. \label{tab:qc_fit}}
\end{table*}

\subsection{Parametrizations used}\label{sec_par_amplitudes}

We use the $K$-matrix parametrization rescaled with the two-body phase space $\rho$ as
\begin{equation}
  K=\rho^{1/2} \hat{K} \rho^{1/2},
\end{equation}
where
\begin{equation}
  \rho=\sqrt{\left(1- \left( \frac{m_\pi +m_N}{\sqrt{s}} \right)^2 \right) \left(1- \left( \frac{m_\pi -m_N}{\sqrt{s}} \right)^2 \right)}.
\end{equation}
The $K$-matrix relates to the phase shifts as
\begin{equation}
  K^{(Jl)}=\tan(\delta_{Jl}).
\end{equation}
As discussed in Sec.~\ref{sec_luscher_analysis}, our analysis includes the phase shift $\delta_{3/2,1}$, where we expect the $\Delta$ resonance that will be quite narrow for our quark masses, and the phase shift $\delta_{1/2,0}$, which is expected to be nonresonant in the energy region considered. We therefore use a Breit-Wigner parametrization for the former,
\begin{equation} \label{eq:BW}
  \hat{K}^{(3/2,1)}=\frac{\sqrt{s} \Gamma(s)}{(m_{\rm BW}^2-s)\rho},
\end{equation}
where $m_{\rm BW}$ denotes the resonance mass and the decay width $\Gamma(s)$ is given by
\begin{equation} \label{eq:gamma_bw}
  \Gamma(s)=\frac{g_{\rm BW}^2}{6\pi}\frac{k^3}{s}
\end{equation}
with the coupling $g_{\rm BW}$, scattering momentum $k$, and center-of-mass energy squared $s$. For the nonresonant $\hat{K}^{(1/2,0)}$ we use the effective-range expansion (ERE) \cite{Landau:1991wop}. We find that working to $0$th-order is sufficient at the level of precision we have, such that
\begin{equation}\label{eq:ere}
  \hat{K}^{(1/2,0)}=\frac{k}{\rho}a_0
\end{equation}
with the $S$-wave scattering length $a_0$.

\subsection{Fit procedure and results}

\begin{figure*}[t]
  \includegraphics[width=0.7\linewidth]{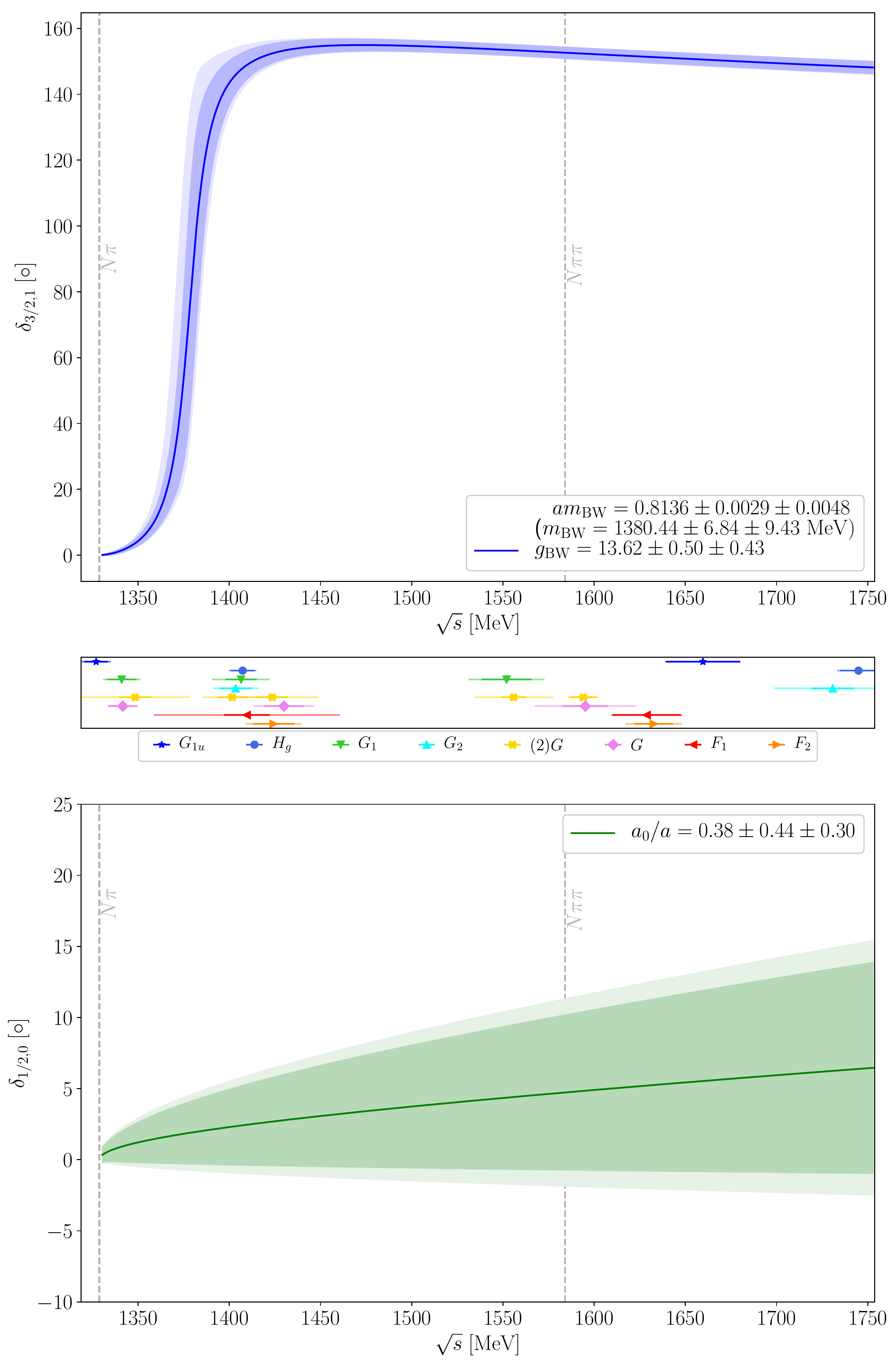}
  \caption{
    Energy-dependence of the $P_{33}$ (upper) and $S_{31}$ (lower) phase shifts from the global fit \textbf{G(a)}. The inner bands show the statistical uncertainty. The outer bands include our estimate of the systematic uncertainty associated with the choice of fit ranges for the two point functions and the selection of energy levels included in the global fit. The center panel shows, with the same axis range, the values of $\sqrt{s}$ for all 21 energy levels included in fit \textbf{G(a)}. The symbols indicate the irreducible representations of these energy levels; the dark error bars show the statistical+scale-setting uncertainties, while the lighter outer error bars also include the estimated systematic uncertainties associated with the fit ranges. }
  \label{fig:PhaseShiftPS}
\end{figure*}

\begin{table*}

  \begin{tabular}{|lllll@{}|}
    \hline
    \rowcolor[gray]{.9}[\tabcolsep] Collaboration             & $m_\pi$ [MeV] & Methodology                    & $m_\Delta$ [MeV]                         & $g_{\Delta\text{-}\pi N}$                                         \\
    \hline
    Verduci 2014 \cite{verduci2014pion}                       & 266(3)        & Distillation, L\"uscher        & $1396(19)_{\rm BW}$                      & 19.90(83)                                                         \\
    Alexandrou et al.~2013 \cite{alexandrou2013determination} & 360           & Michael, McNeile               & 1535(25)                                 & 27.0(0.6)(1.5)                                                    \\
    Alexandrou et al.~2016 \cite{alexandrou2016study}         & 180           & Michael, McNeile               & 1350(50)                                 & 23.7(0.7)(1.1)                                                    \\
    Andersen et al.~2018  \cite{andersen2018elastic}          & 280           & Stoch.~distillation, L\"uscher & $1344(20)_{\rm BW}$                      & 37.1(9.2)                                                         \\
    \hline

    Our result                                                & 255.4(1.6)    & Smeared sources, L\"uscher     & \makecell[l]{$1380(7)(9)_{\rm BW}$,                                                                          \\ $1378(7)(9)_{\rm pole}$ }     & 23.8(2.7)(0.9)                                                         \\
    \hline
    Physical value \cite{PDG2020}                             & 139.5704(2)   & phenomenology, K-matrix        & $1232(1)_{\rm BW}$, $1210(1)_{\rm pole}$ & 29.4(3) \cite{pascalutsa2006chiral}, 28.6(3) \cite{hemmert1995nn} \\
    \hline
  \end{tabular}%

  \caption{Compilation of results for $m_\Delta$ and $g_{\Delta\text{-}\pi N}$. The uncertainties given for the lattice results are statistical/fitting only.}
  \label{tab:comparison}
\end{table*}

Following Ref.~\cite{guo2013coupled} and as in our previous work \cite{Rendon:2020rtw}, we perform a global fit of the model parameters $m_{\rm BW}$, $g_{\rm BW}$, and $a_0$ to all energy levels in all irreps by minimizing the $\chi^2$ function
\begin{align}\label{chi2}
  \chi^2= \sum_{\vec{P},\Lambda,n} \sum_{\vec{P}^\prime,\Lambda^\prime,n^\prime} [C^{-1}]_{\vec{P},\Lambda,n ;\vec{P}^\prime,\Lambda^\prime,n^\prime} \nonumber \\
  \times \left(\sqrt{s_n^{\Lambda,\vec{P}}}^{\text{[data]}} - \sqrt{s_n^{\Lambda,\vec{P}}}^{\text{[model]}} \right) \nonumber                                   \\
  \times \left(\sqrt{s_{n^\prime}^{\Lambda^\prime,\vec{P}^\prime}}^{\text{[data]}} - \sqrt{s_{n^\prime}^{\Lambda^\prime,\vec{P}^\prime}}^{\text{[model]}} \right).
\end{align}
Here, $C$ is the covariance matrix of the energy levels $\sqrt{s_n^{\Lambda,\vec{P}}}^{\text{[data]}}$ measured on the lattice.
The model energies $\sqrt{s_n^{\Lambda,\vec{P}}}^{\text{[model]}}$ are obtained for each parameter guess by finding the roots of the L{\"u}scher quantization conditions (see Table \ref{tab:qc}).
There are 21 energy levels from 8 irreps available for the global fit, as shown in Fig.~\ref{fig:all_irrep_spectra}.

The results for both the global fits and for fits to subsets of energy levels are listed in Table \ref{tab:qc_fit}. Before performing the global fit to all energy levels, we separately considered the irreps that include either only $J=1/2$ or only $J=3/2$ (ignoring $J>3/2$).
The irrep $G_{1u}$ is the only one that contains exclusively $J=1/2$, while there are multiple irreducible representations with exclusively $J=3/2$: $H_g$, $G_2$, $F_1$, and $F_2$.
These initial two fits enable us to obtain a good initial guess for the parameters of the final global fits and assess the stability of the fit over the choice of irreps included.
The fit for the S-wave (labeled \textbf{S}) via irrep $G_{1u}$ is done to only 2 energy levels, resulting in a low $\chi^2/{\rm dof}$.
The other partial fit over irreps containing P-wave only (\textbf{P}) includes 8 energy levels and gives a higher $\chi^2/{\rm dof}$.

For the global fits (\textbf{G}), we implement five different combinations of levels included and choices of $t_{min}/a$ to test the stability of the results and quantify the systematic uncertainty associated with the fits.
The fit to the nominal results for the energy levels from Table \ref{tab:spectrum} is labeled as \textbf{G(a)}, while the fit labeled \textbf{G(a+1)} was done to the energy levels with $t_{min}/a$ increased by one unit throughout.
More focused choices among the noisiest levels are made in the fit \textbf{G(b)}, where we vary $t_{min}/a$ in selected levels based on the results of the stability analysis shown in Figs.~\ref{fig:meff_plot} and \ref{fig:meff_plot_2}. Specifically, this case uses a +1 shift on $t_{min}/a$ on all levels of irreps $G_2,(2)G,F_1,F_2$, the ground state of $G_1$, the first excited of $G$, and +2 on the first excited of $G$. Additionally, we perform the global fit \textbf{G(c)} removing potentially problematic levels from the list in Table \ref{tab:spectrum}: the highest level of irrep $G$ and all levels in irrep $(2)G$.
Furthermore, the global fit labeled \textbf{G(d)} differs from \textbf{G(a)} only by excluding irrep $G_{1u}$.
Overall, we find that the fits provide compatible results and are very stable across several choices.

We select fit \textbf{G(a)} to report the central values and statistical uncertainties of the fit parameters and derived quantities, but then estimate
a systematic uncertainty from the maximum variation in the central value between \textbf{G(a)} and the other four global fits \textbf{G} listed in Table \ref{tab:qc_fit}. That is, for a parameter or derived quantity $y$, we calculate the systematic uncertainty associated with the fit choices as
\begin{equation}\label{eq:systematic}
  \sigma^{sys}_y= \max_i (|y_{\mathbf{G(}i\mathbf{)}}-y_{\mathbf{G(a)}}|) \;\; , \;\; i\in\{ \mathbf{a+1},\mathbf{b},\mathbf{c},\mathbf{d} \}.
\end{equation}
Our final results for the Breit-Wigner parameters and scattering length in lattice units are then
\begin{eqnarray}
\nonumber am_{\rm BW}&=&0.8136 \pm 0.0029 \pm 0.0048,\\
\nonumber g_{\rm BW}&=&13.62\pm0.50 \pm 0.43,\\
 a_0/a&=&0.38 \pm 0.44 \pm 0.30.
\end{eqnarray}
The phase shifts $\delta_{3/2,1}(P_{33})$ and $\delta_{1/2,0}(S_{31})$ from the global fit are plotted as functions of the center-of-mass energy in Fig.~\ref{fig:PhaseShiftPS}. (Recall that a one-to-one mapping of energy levels to scattering phase shifts is not possible in many of the irreps due to the mixing between $J=1/2$ and $J=3/2$. For the irreps without this mixing, we list the results of the direct mapping in Appendix \ref{app:fullphasevssingle}.) Because the $P$-wave phase shift rises rapidly in the region of the resonance, we evaluated separate upper and lower systematic uncertainties that are included in the outer band in Fig.~\ref{fig:PhaseShiftPS}, using the asymmetric generalization of Eq.~(\ref{eq:systematic}) corresponding to the largest shift in each direction.

From the results of the global fit, we also determine the position of the closest $T$-matrix pole in the complex $\sqrt{s}$ plane, associated with the $\Delta$ resonance. Expressing the pole location as $m_\Delta-i \Gamma/2$, we obtain
\begin{eqnarray}
  \nonumber  am_{\Delta}&=&0.8124 \pm 0.0027 \pm 0.0045, \\
  \nonumber  a\Gamma/2&=&0.00484 \pm 0.00061 \pm 0.00084, \\
  \nonumber  m_{\Delta}&=&(1378.3\pm 6.6 \pm 9.0)\:{\rm MeV}, \\
  \Gamma/2&=&(8.2\pm 1.0 \pm 1.4)\:{\rm MeV},
\end{eqnarray}
where the second uncertainty given is the fitting systematic uncertainty estimated using Eq.~(\ref{eq:systematic}). Using our result for $\Gamma$, we then additionally determine the coupling $g_{\Delta\text{-}\pi N}$
from the equation for the decay width in leading-order chiral effective theory
\cite{Jenkins:1990jv,Tang:1996sq,Hemmert:1996xg},
\begin{equation}
  \label{eq:decay}
  \Gamma^{\rm LO}_{\rm EFT}= \frac{ g_{\Delta\text{-}\pi N}^2}{48 \pi}\frac{E_N + m_N}{E_N + E_\pi}\frac{k^{3}}{m_N^2},
\end{equation}
which gives
\begin{equation}
  g_{\Delta\text{-}\pi N}=23.8\pm 2.7 \pm 0.9 .
\end{equation}

The extracted values for the resonance mass $m_\Delta$ and coupling $g_{\Delta\text{-}\pi N}$ are listed with recent results from the literature in Table \ref{tab:comparison}.

Our results for the scattering length $a_0$ are generally consistent with zero within the uncertainties. For the comparison with the literature, we consider the combination $a_0 m_\pi^+$. Our result from global fit \textbf{G(a)} is
\begin{equation}
  a_0 m_{\pi} = 0.057 \pm 0.067 \pm 0.045,
\end{equation}
while the values extracted from experimental data are $-0.0785 \pm 0.0032$ from Ref.~\cite{baru2011precision}, $-0.0894 \pm 0.0017$ from Ref.~\cite{bruns2011chiral} and $-0.101 \pm 0.004$ from Ref.~\cite{hohler1983pion}.

\section{Conclusions}\label{sec_conclusions}

We have presented a determination of elastic nucleon-pion scattering amplitudes for isospin $I=3/2$ using a lattice QCD calculation on a single gauge-field ensemble with pion mass $m_{\pi}\approx 255$ MeV. The baryon $\Delta(1232)$ emerges as the dominant resonance in the $P$-wave with $J^P=3/2^+$ and is the focus of this work. The infinite-volume scattering amplitudes are obtained using the L\"uscher method from the finite-volume energy spectra extracted from correlation matrices built of $\Delta$ and $N\pi$ operators, projected to definite irreducible representations of the lattice symmetry groups. In order to thoroughly map out the energy dependence using just a single volume, it is essential to consider moving frames, where the symmetries are reduced. Many irreps included mix $J=3/2$ and $J=1/2$, and we therefore also extracted the scattering phase shift for the latter.
Each $J$ receives contributions from two values of orbital angular momentum $l$, but at the present level of precision, we can access only a single dominant value of $l$ for each: $l=1$ for $J=\frac32$ and $l=0$ for $J=\frac12$. In addition, we neglect mixing with $J>3/2$.

We performed global fits to the spectra using a Breit-Wigner parametrization for the $P_{33}$ phase shift at energies below the inelastic threshold $N\pi\pi$, and using the leading-order effective-range expansion for the $S_{31}$ phase shift. We also extracted the pole position $m_\Delta-i \Gamma/2$ associated with the $\Delta$ resonance, and the coupling $g_{\Delta\text{-}\pi N}$ that determines the decay width $\Gamma$ at leading order in chiral effective theory. These parameters are listed with other determinations
in Table \ref{tab:comparison}. For our pion mass (and at nonzero lattice spacing), $m_{\Delta}$ is found to be approximately $170$ MeV higher than in nature, while the coupling $g_{\Delta\text{-}\pi N}$ agrees with extractions from experiment at the $2\sigma$ level, given our uncertainties. Our result for the coupling also agrees with previous lattice determinations within the uncertainties. In the $S$ wave, our result for the scattering length is consistent with zero and is also consistent with phenomenological determinations.

Future work will include computations on additional lattice gauge-field ensembles with different spatial volume, which will provide more data points to better constrain the phase shifts extracted and, at the same time, expand on the partial-wave contributions included in the analysis and provide information on remaining finite-volume systematic errors.
Using additional ensembles will also enable us to investigate the dependence on the pion mass and on the lattice spacing. Furthermore, we plan to use the results for the energy levels and scattering amplitudes as inputs to a computation of $N\to N\pi$ electroweak transition matrix elements using formalism of Refs.~\cite{Briceno:2014uqa,Briceno:2015csa}, similarly to what has been done for $\pi\gamma^*\to\pi\pi$ \cite{Briceno:2016kkp,alexandrou2018pi}.

\section{Acknowledgments}\label{sec_acknowledge}
This research used resources of the National Energy Research Scientific Computing Center (NERSC), a U.S. Department of Energy Office of Science User Facility operated under Contract No.~DE-AC02-05CH11231.
SK is supported by the Deutsche Forschungsgemeinschaft grant SFB-TRR 55. SK and GS were partially funded by the IVF of the HGF. LL acknowledges support from the U.S. Department of Energy, Office of Science, through contracts DE-SC0019229 and DE-AC05-06OR23177 (JLAB). SM is supported by the U.S. Department of Energy, Office of Science, Office of High Energy Physics under Award Number DE-SC0009913.
JN and AP acknowledge support by the U.S. Department of Energy, Office of Science, Office of Nuclear Physics under grants DE-SC-0011090 and DE-SC0018121 respectively.
MP gratefully acknowledges support by the Sino-German collaborative research center CRC-110. SP is supported by the Horizon 2020 of the European Commission research and innovation program under the Marie Sklodowska-Curie grant agreement No.~642069. GR is supported by the U.S. Department of Energy, Office of Science, Office of Nuclear Physics, under Contract No.~D{E-S}C0012704 (BNL). SS thanks the RIKEN BNL Research Center for support. AT is supported by the the European Union's Horizon 2020 research and innovation programme under the
Marie Skłodowska Curie European Joint Doctorate STIMULATE, grant No.~765048. We acknowledge the use of the USQCD software QLUA for the calculation of the correlators.

\appendix

\section{One-to-one mapping of energy levels to phase shifts in irreps without mixing between $J=1/2$ and $J=3/2$}
\label{app:fullphasevssingle}

For the irreps that do not mix $J=1/2$ and $J=3/2$, it is possible to directly map individual energy levels to scattering phase shifts using the L\"uscher quantization conditions in Table \ref{tab:qc} (as before, we neglect partial waves higher than $S$ and $P$, respectively). The results of this mapping are shown in Table \ref{tab:phaseshiftscomparison}. For the $G_{1u}$ irrep, the lowest energy level lies just below the $N\pi$ threshold, and we therefore choose to list $ak\cot\delta_{1/2,0}$ (where $k$ is the scattering momentum and $a$ is the lattice spacing) instead of $\delta_{1/2,0}$, as this combination remains real-valued below threshold. For comparison, we also show the phase shifts obtained from the global $K$-matrix fit to all energy levels using the parametrizations (\ref{eq:BW}) and (\ref{eq:ere}). The statistical uncertainties of the energies are propagated to the derived quantities and the systematic uncertainties are computed with Eq.~(\ref{eq:systematic}). For the one-to-one case, the systematic uncertainties are computed as the difference in the values obtained from the fits with $t_{\rm min}$ and $t_{\rm min}+a$.

\begin{table}
  \centering
  \begin{tabular}{| c | c | c | c | c |}
    \hline
    Irrep $\Lambda$ & $n$ & $\sqrt{s}$ [MeV] &\makecell{one-to-one: \\ $\delta_{3/2,1}$ $[^\circ]$ } & \makecell{global: \\ $\delta_{3/2,1}$ $[^\circ]$ }\\
    \hline
    $H_g$           & $1$ & $1407(8)(6)$        & $148(3)(2)$                         & $148(6)(5)$                       \\
    $H_g$           & $2$ & $1745(12)(9)$       & $147(4)(2)$                         & $148(2)(1)$                       \\
    \hline
    $G_2$           & $1$ & $1403(10)(10)$       & $150(4)(3)$                         & $146(8)(6)$                       \\
    $G_2$           & $2$ & $1731(13)(30)$       & $155(11)(23)$                       & $149(2)(1)$                       \\
    \hline
    $F_1$           & $1$ & $1410(13)(50)$       & $131(7)(29)$                        & $149(7)(4)$                       \\
    $F_1$           & $2$ & $1629(20)(7)$        & $142(21)(5)$                        & $151(2)(1)$                       \\
    \hline
    $F_2$           & $1$ & $1424(12)(11)$       & $123(6)(5)$                         & $153(3)(2)$                       \\
    $F_2$           & $2$ & $1633(12)(13)$       & $137(15)(12)$                       & $151(2)(1)$                       \\
    \hline
    \hline
    Irrep $\Lambda$ & $n$ & $\sqrt{s}$ [MeV] &\makecell{one-to-one: \\ $ak\cot\delta_{1/2,0}$}       & \makecell{global: \\ $ak\cot\delta_{1/2,0}$}        \\
    \hline
    $G_{1u}$        & $1$ & $1327(8)(7)$        & $4(15)(3)$                          & $2.6(3.0)(1.2)$                   \\
    $G_{1u}$        & $2$ & $1660(21)(6)$       & $1.4(2.2)(0.1)$                     & $2.6(3.0)(1.2)$                   \\
    \hline
  \end{tabular}
  \caption{Phase shifts obtained using one-to-one mapping of energy levels in irreps that do not mix $J=1/2$ and $J=3/2$, compared to the phase shifts obtained from the global fit at the same center-of-mass energies. For the $G_{1u}$ irrep, where the lowest energy level is found just below the $N\pi$ threshold, we list $ak\cot\delta_{1/2,0}$ instead of $\delta_{1/2,0}$ (where $k$ is the scattering momentum and $a$ is the lattice spacing), because this combination remains real-valued below the threshold.}
  \label{tab:phaseshiftscomparison}
\end{table}

\section{Transformation properties of operators} \label{app:transformation}

In this appendix we list the transformation properties of the momentum-projected field operators under inversions $\mathsf{I}$ and spatial rotations $\mathsf{R}$.
The pseudoscalar pion transforms as
\begin{align} \label{pion_transf}
   & \mathsf{R}\,\pi(\vec{p})\,\mathsf{R}^{-1}=\pi(R\vec{p}) \nonumber \\
   & \mathsf{I}\,\pi(\vec{p})\,\mathsf{I}^{-1}=-\pi(-\vec{p}),
\end{align}
while the nucleon transforms as
\begin{align} \label{nucleon_transf}
   & \mathsf{R}\, N_\alpha(\vec{p})\, \mathsf{R}^{-1}=S (R)_{\alpha\beta}^{-1}  N_\beta(R\vec{p}) \nonumber \\
   & \mathsf{I}\, N_\alpha(\vec{p})\, \mathsf{I}^{-1}=(\gamma_t)_{\alpha\beta} N_\beta(-\vec{p}),
\end{align}
where $S(R)$ is the bi-spinor representation of $SU(2)$. For a rotation of angle $2\pi/n$ around the axis $j$, this is given by
\begin{equation} \label{matrix_S}
  S(R)_{\alpha\beta}=\exp\left( \frac{1}{8}\omega_{\mu\nu} \left[  \gamma_\mu,\gamma_\nu \right]\right)_{\alpha\beta}
\end{equation}
with the antisymmetric tensor $\omega_{kl}=-2\pi \epsilon_{jkl}/n$ and $\omega_{4k}=\omega_{k4}=0$ \cite{morningstar2013extended}.

The vector-spinor Delta operator transforms as
\begin{align} \label{delta_transf}
   & \mathsf{R}\, \Delta_{\alpha k} (\vec{p})\,\mathsf{R}^{-1}=A(R)^{-1}_{kk^\prime} S(R)_{\alpha\beta}^{-1} \Delta_{\beta k^\prime}(R\vec{p}) \nonumber \\
   & \mathsf{I}\,\Delta_{\alpha k}(\vec{p}) \,\mathsf{I}^{-1}=(\gamma_t)_{\alpha\beta}\Delta_{\beta k}(\vec{p})
\end{align}
where $A(R)$ denotes the 3-dimensional $J=1$ irrep of $SU(2)$, and $S(R)$ is given in Eq.~(\ref{matrix_S}).

\onecolumngrid

\section{\texorpdfstring{Matrices $\mathcal{M}_{J l \mu, J' l' \mu'}^{\vec{P}}$}{Matrices M}}
\label{app:matrixM}

Below we provide the matrices $\mathcal{M}_{J l \mu, J' l' \mu'}^{\vec{P}}$ introduced in Eq.~(\ref{eq:mat_Jlmu}), computed for each total momentum $\vec{P}$ including partial wave contributions in $(J=3/2,l=1)$ and $(J=1/2,l=0)$. The momentum labels are given in units of $2\pi/L$.

\begin{footnotesize}
  \begin{align}
    {\cal M}^{(0,0,0)}_{J l \mu, J' l' \mu'} = \bordermatrix{~ &
    \frac{1}{2} 0 \mbox{-}\frac{1}{2}                          & \frac{1}{2} 0 \phantom{\mbox{-}}\frac{1}{2} & \frac{3}{2} 1 \mbox{-}\frac{3}{2} & \frac{3}{2} 1 \mbox{-}\frac{1}{2} & \frac{3}{2} 1 \phantom{\mbox{-}}\frac{1}{2} & \frac{3}{2} 1 \phantom{\mbox{-}}\frac{3}{2} \cr
    \frac{1}{2} 0 \mbox{-}\frac{1}{2}                          & w_{00}                                      & 0                                 & 0                                 & 0                                           & 0                                               & 0 \cr
    \frac{1}{2} 0 \phantom{\mbox{-}}\frac{1}{2}                & 0                                           & w_{00}                            & 0                                 & 0                                           & 0                                               & 0 \cr
    \frac{3}{2} 1 \mbox{-}\frac{3}{2}                          & 0                                           & 0                                 & w_{00}                            & 0                                           & 0                                               & 0 \cr
    \frac{3}{2} 1 \mbox{-}\frac{1}{2}                          & 0                                           & 0                                 & 0                                 & w_{00}                                      & 0                                               & 0 \cr
    \frac{3}{2} 1  \phantom{\mbox{-}}\frac{1}{2}               & 0                                           & 0                                 & 0                                 & 0                                           & w_{00}                                          & 0 \cr
    \frac{3}{2} 1  \phantom{\mbox{-}}\frac{3}{2}               & 0                                           & 0                                 & 0                                 & 0                                           & 0                                               & w_{00} \cr}\
  \end{align}
  \begin{align}
    {\cal M}^{(0,0,1)}_{J l \mu, J' l' \mu'} = \bordermatrix{~ &
    \frac{1}{2} 0 \mbox{-}\frac{1}{2}                          & \frac{1}{2} 0 \phantom{\mbox{-}}\frac{1}{2} & \frac{3}{2} 1 \mbox{-}\frac{3}{2} & \frac{3}{2} 1 \mbox{-}\frac{1}{2} & \frac{3}{2} 1 \phantom{\mbox{-}}\frac{1}{2} & \frac{3}{2} 1 \phantom{\mbox{-}}\frac{3}{2} \cr
    \frac{1}{2} 0 \mbox{-}\frac{1}{2}                          & w_{00}                                      & 0                                 & -i\sqrt{2}w_{10}                  & 0                                           & 0                                               & 0 \cr
    \frac{1}{2} 0 \phantom{\mbox{-}}\frac{1}{2}                & 0                                           & w_{00}                            & 0                                 & 0                                           & -i\sqrt{2}w_{10}                                & 0 \cr
    \frac{3}{2} 1 \mbox{-}\frac{3}{2}                          & 0                                           & 0                                 & w_{00}-w_{20}                     & 0                                           & 0                                               & 0 \cr
    \frac{3}{2} 1 \mbox{-}\frac{1}{2}                          & i\sqrt{2}w_{10}                             & 0                                 & 0                                 & w_{00}+w_{20}                               & 0                                               & 0 \cr
    \frac{3}{2} 1  \phantom{\mbox{-}}\frac{1}{2}               & 0                                           & i\sqrt{2}w_{10}                   & 0                                 & 0                                           & w_{00}+w_{20}                                   & 0 \cr
    \frac{3}{2} 1  \phantom{\mbox{-}}\frac{3}{2}               & 0                                           & 0                                 & 0                                 & 0                                           & 0                                               & w_{00}-w_{20} \cr}\
  \end{align}

  \begin{align}
                                                 & {\cal M}^{(1,1,0)}_{J l \mu, J' l' \mu'} = \cr
                                                 & \bordermatrix{~                                &
    \frac{1}{2} 0 \mbox{-}\frac{1}{2}            & \frac{1}{2} 0 \phantom{\mbox{-}}\frac{1}{2}    & \frac{3}{2} 1 \mbox{-}\frac{3}{2} & \frac{3}{2} 1 \mbox{-}\frac{1}{2} & \frac{3}{2} 1 \phantom{\mbox{-}}\frac{1}{2} & \frac{3}{2} 1 \phantom{\mbox{-}}\frac{3}{2} \cr
    \frac{1}{2} 0 \mbox{-}\frac{1}{2}            & w_{00}                                         & 0                                 & (i-1)\sqrt{3}\text{Re}(w_{11})    & 0                                           & (1+i)\text{Re}(w_{11})                          & 0 \cr
    \frac{1}{2} 0 \phantom{\mbox{-}}\frac{1}{2}  & 0                                              & w_{00}                            & 0                                 & (i-1)\text{Re}(w_{11})                      & 0                                               & (1+i)\sqrt{3}\text{Re}(w_{11}) \cr
    \frac{3}{2} 1 \mbox{-}\frac{3}{2}            & (1+i)\sqrt{3}\text{Re}(w_{11})                 & 0                                 & w_{00}-w_{20}                     & 0                                           & \sqrt{2}w_{22}                                  & 0 \cr
    \frac{3}{2} 1 \mbox{-}\frac{1}{2}            & 0                                              & (1+i)\text{Re}(w_{11})            & 0                                 & w_{00}+w_{20}                               & 0                                               & \sqrt{2}w_{22} \cr
    \frac{3}{2} 1  \phantom{\mbox{-}}\frac{1}{2} & (i-1)\text{Re}(w_{11})                         & 0                                 & -\sqrt{2}w_{22}                   & 0                                           & w_{00}+w_{20}                                   & 0 \cr
    \frac{3}{2} 1  \phantom{\mbox{-}}\frac{3}{2} & 0                                              & (i-1)\sqrt{3}\text{Re}(w_{11})    & 0                                 & -\sqrt{2}w_{22}                             & 0                                               & w_{00}-w_{20} \cr}\ \cr
  \end{align}

  \begin{align}
    {\cal M}^{(1,1,1)}_{J l \mu, J' l' \mu'} = \bordermatrix{~ &
    \frac{1}{2} 0 \mbox{-}\frac{1}{2}                          & \frac{1}{2} 0 \phantom{\mbox{-}}\frac{1}{2} & \frac{3}{2} 1 \mbox{-}\frac{3}{2} & \frac{3}{2} 1 \mbox{-}\frac{1}{2} & \frac{3}{2} 1 \phantom{\mbox{-}}\frac{1}{2} & \frac{3}{2} 1 \phantom{\mbox{-}}\frac{3}{2} \cr
    \frac{1}{2} 0 \mbox{-}\frac{1}{2}                          & w_{00}                                      & 0                                 & (1-i)\sqrt{\frac{3}{2}}w_{10}     & -i\sqrt{2}w_{10}                            & -i e^{\frac{i\pi}{4}}w_{10}                     & 0 \cr
    \frac{1}{2} 0 \phantom{\mbox{-}}\frac{1}{2}                & 0                                           & w_{00}                            & 0                                 & 0                                           & -i\sqrt{2}w_{10}                                & (1+i)\sqrt{\frac{3}{2}}w_{10} \cr
    \frac{3}{2} 1 \mbox{-}\frac{3}{2}                          & (1+i)\sqrt{\frac{3}{2}}w_{10}               & 0                                 & w_{00}                            & -2 e^{\frac{i\pi}{4}} w_{22}                & \sqrt{2}w_{22}                                  & 0 \cr
    \frac{3}{2} 1 \mbox{-}\frac{1}{2}                          & i\sqrt{2}w_{10}                             & i e^{\frac{i\pi}{4}}w_{10}        & 2 e^{\frac{i\pi}{4}} w_{22}       & w_{00}                                      & 0                                               & \sqrt{2}w_{22} \cr
    \frac{3}{2} 1  \phantom{\mbox{-}}\frac{1}{2}               & -i e^{\frac{i\pi}{4}}w_{10}                 & i\sqrt{2}w_{10}                   & -\sqrt{2}w_{22}                   & 0                                           & w_{00}                                          & 2 e^{\frac{i\pi}{4}} w_{22} \cr
    \frac{3}{2} 1  \phantom{\mbox{-}}\frac{3}{2}               & 0                                           & (1-i)\sqrt{\frac{3}{2}}w_{10}     & 0                                 & -\sqrt{2}w_{22}                             & -2 e^{\frac{i\pi}{4}} w_{22}                    & w_{00} \cr}\
  \end{align}

\end{footnotesize}

\twocolumngrid

\bibliography{bibliography}

\end{document}